\documentclass[preprint,11pt]{aastex}
\shorttitle{Initial Mass Function. I.}
\shortauthors{PARRAVANO, McKEE AND HOLLENBACH}

\begin{document}

\title{An Initial Mass Function for Individual Stars in Galactic Disks: I. Constraining the Shape of the IMF}

\author{Antonio Parravano\altaffilmark{1}, 
Christopher F. McKee\altaffilmark{2},  and
David J. Hollenbach\altaffilmark{3,4}}

\altaffiltext{1}{Universidad de Los Andes, Centro de F\'{\i}sica
Fundamental, M\'erida 5101a, Venezuela}

\altaffiltext{2}{Physics Department and Astronomy Department
University of California at Berkeley, Berkeley, CA 94720} 

\altaffiltext{3}{NASA Ames Research Center, MS 245-3, Moffett Field,
 CA 94035}

\altaffiltext{4}{SETI Institute, 515 N. Whisman Road, Mountain View,
 CA 94043}

\newcommand{\beq}	{\begin{equation}}
\newcommand{\eeq}	{\end{equation}}
\newcommand{\beqa}{\begin{eqnarray}}
\newcommand{\eeqa}{\end{eqnarray}}
\newcommand{\avg}[1]  {{\langle #1 \rangle}} 
\newcommand{\dis}{\displaystyle}
\newcommand{\e}	{$^{-1}$}
\newcommand{\ee}	{$^{-2}$}
\newcommand{\eee}	{$^{-3}$}

\newcommand{\calm}	{{\cal M}}
\newcommand{\caln}	{{\cal N}}
\newcommand{\calr}	{{\cal R}}
\newcommand{\pbyp}[1]	{{{\partial\hfil}\over{\partial#1}}}
\newcommand{\ppbyp}[2]	{{{\partial#1}\over{\partial#2}}}

\def\bd{{\rm bd}}
\def\fnbd{F_{n,\,\rm bd}}
\def\hc{{\rm HC}}
\def\imfst{IMF$_{\rm STPL}$}
\def\nsh{{\cal N}_{*h}}
\def\dns{\dot{\cal N}_{*}}
\def\dnst{\dot{\cal N}_{*T}}
\def\dnsms{\dot{\cal N}_{*,\,\rm ms}}
\def\dnsh{\dot {\cal N}_{*h}}
\def\nsm{\caln_*(m)}
\def\dnsm{\dot\caln_*(m)}
\def\mbd{m_{\rm bd}}
\def\mch{m_{\rm ch}}
\def\mpk{m_{\rm peak}}
\def\mtil{\tilde m}
\def\muh{\mu_h}
\def\peak{{\rm peak}}
\def\pn{{\rm PN}}
\def\Rtil{\tilde R}
\def\stpl{{\rm STPL}}
\def\nst{\varsigma _{*t}}
\def\dsst{\dot \varsigma _{*t}}
\def\ssh{\varsigma _{*h}}
\def\dssh{\dot \varsigma _{*h}}
\def\ssm{\varsigma _{*}(m)}

\def\eff			{{{\rm eff}}}
\def\ms                  {{\rm ms}}
\def\meff                {m_{\rm \eff}}

\def\ednb                {\dot\caln_b}
\def\ednbo               {\dot\caln_{b,\, 1}}
\def\ednbt               {\dot\caln_{b,\, T}}
\def\enbo                {\caln_{b,\, 1}}
\def\enbt                {\caln_{b,\, T}}
\def\ednbe               {\dot\caln_b^\eff}
\def\enbe                {\caln_b^\eff}
\def\ednbeo              {\dot\caln_{b,\,1}^\eff}
\def\ednbet              {\dot\caln_{b,\,T}^\eff}

\def\ind			{{\rm ind}}
\def\sys			{{\rm sys}}
\def\bd			{{\rm bd}}
\def\sup			{{\rm sup}}
\def\nhi 		{{\caln_{h}^\ind}}
\def\nhe		{{\caln_{h}^\eff}}
\def\nli 		{{\caln_{*\ell}^\ind}}
\def\nle		{{\caln_{*\ell}^\eff}}
\def\nbdhi 		{{\caln_{\bd}^\ind}}
\def\nbdhe  		{{\caln_{\bd}^\eff}}
\def\sl			{{s,\,\ell}}
\def\sh			{{s,\,h}}
\def\bl			{{b,\,\ell}}
\def\bh			{{b,\,h}}
\def\bdh			{{\bd}}
\def\ftl			{\phi_{2\ell}}
\def\ftbd			{\phi_{2\bd}}
\def\pb			{{\rm pb}}
\def\jpb			{{j_\pb}}
\def\ltsimeq{\,\raise 0.3 ex\hbox{$ < $}\kern -0.75 em
 \lower 0.7 ex\hbox{$\sim$}\,}
\def\gtsimeq{\,\raise 0.3 ex\hbox{$ > $}\kern -0.75 em
 \lower 0.7 ex\hbox{$\sim$}\,}

\def\half {\hbox{$1 \over 2$}}
\let\ga=\gtsimeq
\let\la=\ltsimeq
\def\beq{\begin{equation}}
\def\eeq{\end{equation}}
\def\avg #1{\langle #1\rangle}
\def\calm{{\cal M}}
\def\caln{{\cal N}}
\def\calr{{\cal R}}
\def\dis{\displaystyle}
\def\vecr{{\bf r}}

\def\bd{{\rm bd}}
\def\MK{{\rm MK}}
\def\fnbd{F_{n,\,\rm bd}}
\def\hc{{\rm HC}}
\def\imfst{IMF$_{\rm STPL}$}
\def\nsh{{\cal N}_{*h}}
\def\dns{\dot{\cal N}_{*}}
\def\dnst{\dot{\cal N}_{*T}}
\def\dnsms{\dot{\cal N}_{*,\,\rm ms}}
\def\dnsh{\dot {\cal N}_{*h}}
\def\nsm{\caln_*(m)}
\def\dnsm{\dot\caln_*(m)}
\def\mbd{m_{\rm bd}}
\def\mch{m_{\rm ch}}
\def\mpk{m_{\rm peak}}
\def\mtil{\tilde m}
\def\muh{\mu_h}
\def\pn{{\rm PN}}
\def\Rtil{\tilde R}
\def\stpl{{\rm STPL}}
\def\nst{\varsigma _{*t}}
\def\dsst{\dot \varsigma _{*t}}
\def\ssh{\varsigma _{*h}}
\def\dssh{\dot \varsigma _{*h}}
\def\ssm{\varsigma _{*}(m)}

\def\iint                {\int\hspace{-7pt}\int}
\def\eff                 {{\rm eff}}
\def\aeff                {A_\eff}
\def\leff                {L_\eff}
\def\lveff               {L_{V,\eff}}
\def\ind		     {{\rm ind}}
\def\ms                  {{\rm ms}}
\def\meff                {m_{\rm \eff}}
\def\pk			{{\rm peak}}
\def\sys			{{\rm sys}}

\def\imffp		 {IMF$_{4p}$}
\def\jb                  {{j_{b0}}}
\def\jv                  {{j_V}}
\def\jt                  {{j_T}}

\def\Nu{{\cal N}_{*h,u}}
\def\Nl{{\cal N}_{*h,l}}
\def\Nt{{\cal N}_{*h,T}}
\def\Non{{\cal N}_{*h,1}}
\def\Noncr{{\cal N}_{*h,1cr}}
\def\Ntw{{\cal N}_{*h,2}}
\def\Ntwcr{{\cal N}_{*h,2cr}}
\def\Nth{{\cal N}_{*h,3}}
\def\d{n_{ac}(\Nh)}
\def\dd{n_a(\Nh)}
\def\ddd{{\dot n}_{ac}(\Nh)}
\def\dddd{{\dot n}_{a}(\Nh)}
\def\sd{\varsigma _a(\Nh)}
\def\sdd{{\dot {\varsigma}} _a(\Nh)}
\def\s0{{\dot {\varsigma}} _{a0}}
\def\Ut{U_{\rm band}}
\def\Us{U_s}
\def\Ub{U_{\rm back}(\Us)}
\def\Um{<\Ut >}
\def\Umed{U_{\rm med}}
\def\Umeds{U_{{\rm med},s}}
\def\Uz{U_0}
\def\Uo{U_1}
\def\Uoa{U_{1a}}
\def\Utw{U_2}
\def\Uth{U_3}
\def\del{\Delta _{\rm band}}
\def\H{\sqrt {\pi}H_*}
\def\ph{\Phi _s(\Us)}
\def\Ps{P_s(>\Ut)}
\def\Pss{P_s(>\Us)}
\def\Pt{P_t(>\Ut)}
\def\ps{p_s(\Ut)}
\def\pss{p_s(\Us)}
\def\rn{r_{\Nh}(\Us)}
\def\rco{r_1}
\def\rcoa{r_{1a}}
\def\rct{r_2}
\def\Lf{L_{\rm band}(\Nh)}
\def\L0{L_{0,\rm band}}
\def\Lfuv{L_{0,\rm FUV}}
\def\lm{\ell _{\rm mfp}(>\Ut)}
\def\ta{t_U(>\Ut )}
\def\t0{t_{\rm band}}

\def\esse{\Sigma _{*,\rm ever}}
\def\pcsq{pc$^{-2}$}
\def\epcsq{\rm pc^{-2}}
\def\emsun{M_\odot}
\def\msun{$M_\odot \,$}
\def\IMFours{IMF$_{\rm GAL}$}
\def\PDMFours{PDMF$_{\rm GAL}$}

\def\embd		 {m_{\rm bd}}
\def\empl		 {m_{\rm pl}}
\def\edNso		 {\dot\caln_{*1}}
\def\edNst		 {\dot\caln_{*T}}
\def\edsso		 {\dot\varsigma_{*1}}
\def\edsst		 {\dot\varsigma_{*T}}
\def\emch		 {m_{\rm ch}}
\def\emcheff		 {m_{\rm ch,\, eff}}
\def\emchind		 {m_{\rm ch,\, ind}}
\def\embch		 {m_{b,\, \rm ch}}

\def\ednso		 {\dot\caln_{*1}}
\def\ednst		 {\dot\caln_{*T}}

\def\emuh{\mu_h}
\def\eNsm{\caln_*(m)}
\def\eNsh{\caln_{*h}}
\def\eNshtyp{\caln_{*h,\, \rm typ}}
\def\eNshtypmax{\caln_{*h,\, \rm typ\; max}}
\def\eNshl{{\caln_{*h,l}}}
\def\eNshu{{\caln_{*h,u}}}
\def\ess{\varsigma _{*}}
\def\essSTPL{\varsigma _{*,{\rm STPL}}}
\def\essh{\varsigma _{*h}}
\def\essm{\varsigma _{*}(m)}
\def\edNsm{\dot\caln_*(m)}
\def\edNsmt{\dot\caln_*(m,t)}
\def\eNa{{\cal N}_a}
\def\edsa{\dot \varsigma _{a}}
\def\edss{\dot \varsigma _{*}}
\def\edssh{\dot \varsigma _{*h}}
\def\edsso{\dot \varsigma _{1}}
\def\edsst{\dot \varsigma _{*T}}
\def\edsshtyp{\dot \varsigma _{*h,\, \rm typ}}
\def\edssm{\dot \varsigma _{*}(m)}
\def\edNsh{\dot {\cal N}_{*h}}
\def\edNshT{\dot {\cal N}_{*h}}
\def\edNs{\dot {\cal N}_{*}}
\def\eavtion{\avg{t_{\rm ion}}}
\def\dens{\dot \varsigma _{*t}}
\def\etsim{{t_{\rm sim}}}
\def\ednb                {\dot\caln_b}
\def\ednbo               {\dot\caln_{b,\, 1}}
\def\ednbt               {\dot\caln_{b,\, T}}
\def\enbo                {\caln_{b,\, 1}}
\def\enbt                {\caln_{b,\, T}}
\def\ednbe               {\dot\caln_b^\eff}
\def\enbe                {\caln_b^\eff}
\def\ednbeo              {\dot\caln_{b,\,1}^\eff}
\def\ednbet              {\dot\caln_{b,\,T}^\eff}
\def\edst                {\dot \varsigma_T}
\def\edsb                {\dot \varsigma_b}
\def\edsbo               {\dot \varsigma_{b,\, 1}}
\def\edsbt               {\dot \varsigma_{b,\, T}}
\def\edsbe               {\dot \varsigma_b^\eff}
\def\edsboe              {\edsbo^\eff}

\begin{abstract}

We derive a semi-empirical galactic initial mass function (IMF) 
from observational constraints.
We assume that the star formation rate in a galaxy can be expressed 
as the product of
the IMF, $\psi (m)$,   which is a smooth function of mass $m$ 
(in units of \msun),
and a time- and space-dependent total rate of star formation per unit area  of galactic disk, $\edsst$. 
The mass dependence of the
proposed IMF is determined by
five parameters: the low-mass slope $\gamma$, the high-mass slope
$-\Gamma$
(taken to be the Salpeter value, -1.35), 
the characteristic mass $\emch$ (which is close to the mass $m_{\rm peak}$ at which the IMF turns
over), and the lower and upper limits on the mass, 
$m_\ell$ (taken to be 0.004) and $m_u$ (taken to be 120).
The star formation rate in terms of number of stars per unit area of
galactic disk per unit logarithmic mass interval, is proportional to the IMF:
$$
\edss (m) \equiv \frac{d^2\edNsm}{dA\; d\ln m}
	   \equiv\edsst \psi(m)= C
	   \edsst \, m^{-\Gamma} 
	   \left\{1-\exp\left[{-(m/\emch)^{\gamma +
	   \Gamma}}\right]\right\},
$$
where $\cal N_*$ is the number of stars, $m_\ell<m<m_u$ is the range of stellar masses.
The values of $\gamma$ and $\emch$ are derived from two integral constraints: i) the 
ratio of the number density of stars in the range $m=0.1-0.6$ to
that in the range $m=0.6-0.8$ 
as inferred from the mass distribution of field stars
in the local neighborhood, and ii) 
the ratio of the number of stars in the range $m=0.08 - 1$
to the number of
brown dwarfs
in the range $m=0.03-0.08$ in young clusters.
The IMF satisfying the above constraints is characterized by the parameters 
$\gamma=0.51$ and
$\emch=0.35$ (which corresponds to $m_{\rm peak}=0.27$). 
This IMF agrees quite well with the Chabrier (2005) IMF for the entire
mass range over which we have compared with data, but predicts significantly more stars
with masses $< 0.03\, M_\odot$;
we also compare with other IMFs in current use.
We give a number of important parameters implied by the IMF, such as the 
fractional number
of brown dwarfs and high-mass stars formed at a given time, the average mass of 
a newly-formed star, and the mass of
stars formed per high-mass star.

\end{abstract}


\section{Introduction}

The Initial Mass Function (IMF) is a fundamental ingredient for 
modeling any system containing stars -- from clusters to galaxies to
the luminous universe. 
Since the pioneering work by Salpeter (1955), the IMF has been
derived for a variety of systems, such as clusters of different ages, 
field stars,
the galactic bulge, globular clusters, and nearby galaxies.
Comprehensive studies by
Miller \& Scalo (1979), Scalo (1986), Kroupa (2001), and Chabrier (2003b, 2005) have provided
the community with standard IMFs that
have allowed the construction of innumerable models with a common basis.
A comprehensive review of the current state of IMF studies has been given by
Bastian, Covey \& Meyer (2010). Earlier 
reviews of the theoretical and observational studies of the IMF
are in Scalo (1998a) and (1998b) respectively.
It is impressive that after more half a century, the Salpeter
work is still the accepted IMF for intermediate and high-mass stars.
At low masses the situation has evolved much more. The ever increasing
sensitivity and resolution of observations, together with the theoretical
means to infer the stellar masses from the observable quantities,
have resulted in the discovery of a turn-over in the IMF at 
$\sim 0.5$ \msun and
in a progressive reduction of the errors in the low-mass IMF.
Only few years ago the existence of a flattening or
a turnover in the IMF at subsolar masses was controversial,
but now its existence is well established. However, uncertainties in the IMF grow 
as mass decreases
below the hydrogen-burning mass limit ($\sim 0.08$ \msun) into 
the realm of brown dwarfs.

	  One of the problems that plagues attempts to
determine the IMF is that of unresolved binary (or multiple)
star systems.  One can distinguish three different IMFs:

\begin{itemize}

\item{} the {\it individual-star IMF, or ``true'' IMF}, $\psi(m)$, 
in which each
star that is a member of a multiple system is counted
separately; 

\item{} the {\it system IMF}, $\psi^\sys(m)$, in which each multiple stellar
system and each single star
is counted as one object with its total mass; and

\item{} the {\it effective system IMF} (or {\it effective IMF} for short), $\psi^\eff(m)$,
in which unresolved
binaries are counted as single stars with an effective mass
$\meff$. Note that Chabrier (2003b) adopted a different terminology: he
used the term ``system IMF'' for $\psi^\eff$ and did not consider what
we define as $\psi^\sys$. 

\end{itemize}
The relation between the effective mass and the system
mass depends on how the mass is determined. For stars with known distances
(such as those in clusters), the absolute magnitude of the system is observed
and the effective mass is inferred from a mass-magnitude relation.
For field stars that do
not have independent distance determinations, a color-magnitude relation
is used to infer the absolute magnitude of the system and then a mass-magnitude
relation is used to infer the effective mass. Interestingly,
Gould, Bahcall, \& Flynn (1996) showed that the stellar
mass per unit area of Galactic disk
inferred from the effective IMF agrees reasonably
well with that from the true IMF.
It should be noted that theories of the IMF, such as those of Padoan \& Nordlund (2002;
see also Padoan et al. 2007) and
of Hennebelle \& Chabrier (2008, 2009) predict the system IMF, whereas observations
that cannot resolve close binaries determine an effective IMF.

	The goal of this paper is to infer the individual-star
IMF for stars in the disk of the Galaxy. In this
paper, we shall use the term ``star" to
refer to star-like objects, whether they burn
nuclear fuel or not, as well as stellar remnants.
Thus, brown dwarfs, white dwarfs, neutron stars and stellar-mass black holes
(which are sometimes referred to as "collapsed stars")
are included in the term "star." 
The terminology for objects below the hydrogen-burning
limit has not been fixed yet (e.g., see Basri 2003, Chabrier et al. 2005); 
we use the term "brown dwarfs" for all stars below the hydrogen-burning limit.
Evolved objects such as white dwarfs
do not enter the IMF; the IMF thus covers  
brown dwarfs and main-sequence (i.e., hydrogen-burning) stars.
We distinguish stars from planets, which form in 
circumstellar disks and have an excess of heavy elements 
(Chabrier et al. 2007, Whitworth et al. 2007).
In principle, free-floating planets could be confused observationally
with brown  dwarfs (e.g., Liu \& Butler 2002);
based on observations of extrasolar planets, such free-floating
planets are expected to have masses less than 10 Jupiter masses.

In addressing this
problem, we assume that {\it the IMF for stars in galactic disks is
both universal and simple:}

{\it Universality:} Although significant variations in the IMF
are observed in the disk of the Galaxy (e.g., Scalo 1998a), for the
most part these are consistent with random sampling from
a universal IMF (Elmegreen 1997, 1999; Kroupa 2002; Bastian et al. 2010). 
Fluctuations about the mean IMF can arise in IMF models based on deterministic
chaos (S\'anchez \& Parravano 1999).
Evidence for evolution of the IMF with redshift is summarized by Elmegreen (2009),
but none of this evidence deals specifically with disk galaxies.
Treu et al. (2010) find that for a sample of early-type galaxies a Salpeter IMF 
provides light-to-mass ratios more consistent with observation
than does a Chabrier (2003b) IMF, and conclude that massive
early-type galaxies cannot have both a universal IMF and a universal dark matter halo.
Again, this result does not deal with galactic disks.
Observational evidence that the upper part of the IMF varies with surface brightness is
inconclusive: Meurer et al. (2009)
found that the ratio of H$\alpha$ to FUV increased
with surface brightness,
which could be explained if massive stars form only in massive clusters (e.g., Pflamm-Altenburg et al. 2009)
or if the formation of massive stars requires high gas surface densities (Krumholz \& McKee 2008).
On the other hand, detailed observations of outer regions in Arp 78 (Kotulla et al. 2008) and M81 
(Gogarten et al. 2009) indicate that the high-mass IMF is normal in these regions.
In any case,
most star formation occurs in the disk regions where massive stars also form,
and we assume that, when averaged over large regions of
these parts of galactic disks and over 
long intervals of time, the mass distribution of stars is drawn from
a universal IMF, $\psi(m)$. 
This is consistent with the conclusion of Bastian et al. (2010), who 
conclude that there is no clear evidence that the IMF varies strongly and systematically as a function of initial conditions after the first few generations of stars.

{\it Simplicity.} We further assume that
the IMF is a smooth function of stellar mass that
is characterized by a relatively small number of parameters.
There are no known physical processes that would make the
IMF complex, particularly after averaging over
the disparate physical conditions in Galactic star-forming
regions. The molecular clouds in which stars form are
highly turbulent (see Vazquez-Semadeni et al. 2000 and McKee \& Ostriker 2007
for reviews), and the
only scales that have been identified observationally
in these clouds
are the thermal Jeans mass and the mass of the entire
molecular cloud (Williams, Blitz, \& McKee 2000).
The close correspondence between the mass function of
molecular cores and the IMF (reviewed in McKee \& Ostriker 2007)
suggests that this lack of scale should carry over from
molecular clouds to stellar systems.
There are several ways in which complexity might enter into the individual-star 
IMF: the stellar mass is only a fraction of the mass of the core from which
it forms, the rest being ejected by protostellar outflows and, for massive
stars, by photoevaporation; the formation of binaries, etc., is
due to fragmentation of the core; and finally, the multiplicity of stellar
systems evolves with time. In each case, the associated physical processes
could in principle leave their mark on the IMF. We assume that these effects
are not large, and in any case they tend to be averaged out by considering the
total population of stars from many different star-forming regions.

A significant departure from simplicity would occur if there is more than one distinct star
formation mechanism. Indeed, several mechanisms have been suggested for
the formation of very low-mass stars (VLMSs) and brown dwarfs (see the review by
Whitworth et al. 2007). Thies and Kroupa (2007) have argued that
observations of the distribution of the distances of separation of binaries supports the existence of
a separate mechanism for the formation of brown dwarfs and VLMSs. As they
point out, there is no reason for the star formation mechanism
to depend on the hydrogen burning limit, so that there should be a significant 
mass range in which the two mechanisms are both operative. In that case, the
total IMF would presumably still be smooth and relatively simple. 

Contrary to our assumption of simplicity, some
observations do show features in the inferred
IMF; for example, the mass function of nearby stars measured
by Reid et al (2002) has upticks both at $m\simeq 0.1$
and at $m\simeq 1.0$ (throughout this paper
$m$ is the stellar mass in units of solar masses). We assume that these
apparent features are due to systematic effects such
as uncertainties in the mass/magnitude relation 
and the dependence of this relation on metallicity.

	Based on the assumption that the IMF is simple, we are free to
adopt a simple analytic expression for it. Our
approach in this paper 
is thus quite different from the classical one of
inferring the IMF at each mass directly from the data
(e.g., Scalo 1986).  
Furthermore, since our form for the IMF is simple, we use integral 
constraints that are observationally relatively well determined to
infer the parameters of the IMF.
The specific functional form for the IMF that we
adopt is
\beq
\psi(m)\propto m^{-\Gamma}\{1-\exp[-(m/\mch)^{\gamma+\Gamma}]\}.
\label{eq:psi1}
\eeq
which, as we shall show below, is consistent with existing observations.
We assume that equation (\ref{eq:psi1}) applies
to the mass range $m_\ell \leq m \leq m_u$, and that $\psi=0$ otherwise;
thus our assumed form for the IMF has the form of two smoothly joined,
cut-off power laws.
Since this function 
approaches a power law at both low stellar masses ($\psi\propto m^{\gamma}$)
and at high stellar masses ($\psi\propto m^{-\Gamma}$), we term it
the Smoothed Two-Power Law (STPL) form for the IMF.
 It has a total of 5 parameters: the two power laws, the characteristic mass,
$\mch$, and the upper and lower limits on the mass, $m_\ell$ and $m_u$.

Historically, this functional form was first proposed by Paresce \& De Marchi (2000)
to describe the Present Day Mass Function (PDMF) of globular clusters. 
(They termed this form of the mass function the Tapered Power Law,
but we prefer to emphasize that there are two power laws involved.)
De Marchi \& Paresce (2001) showed that this form described the PDMFs of
galactic clusters as well, and that in such clusters the characteristic mass increases
with age. De Marchi, Paresce \& Portegies Zwart (2003) suggested that the
PDMF for field stars could be built up by a superposition of STPL functions 
with different values of $\mch$.
At a conference in 2003, Parravano, McKee \& Hollenbach
(2006) proposed the STPL form for the IMF;
Hollenbach, Parravano \& McKee (2005)
presented an updated version of this form of the STPL form for the IMF.
A similar form (but with $\gamma=0$) has been used 
by Elmegreen (2006) and Elmegreen et al. (2008).
De Marchi, Paresce \& Portegies Zwart (2010) have given a detailed discussion
of fitting STPLs to the mass functions of a  number of young galactic clusters 
(ages less than 1 Gyr) and globular clusters
over the mass range $(0.1-10) M_\odot$. For the young galactic clusters,
they find $\Gamma\simeq 1$ and $\gamma\simeq 1.5$. The characteristic mass
increases with age, and they infer an initial value of $\mch=0.15$.

Our approach toward inferring the IMF is complementary to that of De Marchi et al.
First, we consider a broader mass range, using data from $0.03 M_\odot$ to
over $100 M_\odot$; the inclusion of higher mass stars leads to a somewhat larger
value of $\Gamma$ than they found. Second, we rely either on very young clusters
(to infer the brown dwarf fraction) or observations of local field stars (to fix the
shape of the IMF for low-mass main sequence stars). The local field stars are
the result of the dissolution of a very large number of galactic clusters
of the type analyzed by De Marchi et al. 
In our approach, we infer the IMF of intermediate-mass stars by 
interpolation between
the IMF of subsolar objects and the IMF of massive stars,
thereby avoiding distortions of the 
PDMF due to time variations
in the star formation rate over the lifetime of 
intermediate mass stars.
Our approach is similar to that of Kroupa (2001, 2002),
who introduced three and four power-law fits to the IMF;
our form is simpler in that it has one characteristic
mass instead of two or three and is more amenable to analytic computation.

It is customary to fit the low-mass part of the IMF with a log-normal form
(e.g., Miller \& Scalo 1979; Chabrier 2003b, 2005), which is consistent
with the probability distribution functions
(PDFs) for the density found in simulations of supersonically turbulent,
non-self-gravitating molecular clouds
(e.g., Nordlund \& Padoan 1999; Ostriker, Stone, \& Gammie 2001).
However, as first shown by Passot \& Vazquez-Semadeni (1998), an equation of
state with
an adiabatic index less than unity leads to a power-law tail in the density PDF at
high densities,
which would increase the rate of production of brown dwarfs compared to a log-normal
PDF (Hennebelle \& Chabrier 2009).
Additional physical effects, such as 
self-gravity (e.g., Li et al. 2003; Bonnell et al. 2008)
and protostellar outflows (Li \& Nakamura 2006), also alter the properties of the turbulence. 
Furthermore, the low-mass portion of the mass spectrum 
of density fluctuations is very difficult
to determine computationally due to limitations in resolution.
Thus, at the present time, theory offers little guidance on the actual form of the 
low-mass portion of the IMF, and it is an important observational goal to determine this.
Bastian et al. (2010) point out that if the mid-range of stellar masses
is described by a log-normal form and if there are power-law tails at both
high and low masses, then a total of 8 parameters is needed to describe the
IMF, which is considerably more complicated than the STPL form.
As shown in \S \ref{sec:comparison}, the STPL form of the IMF determined here
agrees with the Chabrier (2005) form, which is log-normal below $1 M_\odot$, 
to within 30\% for $m>0.03$; the IMFs differ by more than a factor 2 only for $m<0.017$.
An observational determination of whether the log-normal form extends to
the lowest mass stars will be challenging,
but it is important in order to understand the formation of these stars.

In this paper (Paper I) we use three constraints to calibrate the STPL form of the IMF:
(i) the slope of the high-mass IMF, $-\Gamma$; 
(ii) the ratio $R_{MK}$ of the number of stars (mainly M dwarfs) in the range
$0.1-0.6 M_\odot$ to the number 
of stars (mainly K dwarfs) in the range $0.6-0.8 M_\odot$; 
and (iii) the ratio $R_\bd$ 
of the number of stars in the mass range $0.08 - 1 M_\odot$ 
divided by the number of objects between 
$0.03$ and $0.08 M_\odot$.
In order to test the accuracy of the STPL form of the IMF, 
as well as those proposed by Kroupa (2002), Padoan \& Nordlund (2002), Chabrier (2005), and
Hennebelle \& Chabrier (2008),
we compare the IMF shapes with
the observations summarized in this paper. 

In Paper II (Parravano, McKee, \& Hollenbach 2010) we carry out tests that depend
on the normalization of the IMF
(i.e., surface density of stars in the Galactic disk) as well as on the shape. 
These tests include the local PDMF, the
abundance of white dwarfs and red giants, and the value of the surface density of all stars at
the solar circle.
We then infer the current rate of star formation from observations of the ionizing photon luminosity from massive stars in the Galaxy; this rate is an average
over the last $\sim 5$ Myr. We compare this current rate with the time-averaged rate
 over the lifetime of the Galaxy ($\sim 11$ Gyr), as determined from observations of the current
 surface density of low-mass stars that have survived this entire period.

This paper is organized as follows.
In \S 2 we define the basic quantities needed to describe the IMF and
PDMF and we discuss several different proposed forms for the IMF.
In \S 3, we use observational data to determine the shape of the IMF.
This shape is compared with observation in \S 4,
and finally, in \S 5 we summarize the results.

\section{Forms for the IMF}

\subsection{Basic Definitions}
\label{sec:basic}

Let $d\eNsm$
be the number of main-sequence stars and 
brown dwarfs with masses between $m$ and $m+dm$. For
stars above the hydrogen-burning limit
($\embd\simeq 0.075$---Burrows et al 2001),
we take $m$ to be the Zero Age Main Sequence (ZAMS) mass.
We express the PDMF of the main-sequence stars and brown dwarfs 
as $d\caln_*/d\ln m$, 
which is simply the number of such stars in a logarithmic mass
interval.  We denote the PDMF per unit area of Galactic disk by 
\beq
\varsigma_*(m)\equiv \frac{d^2\caln_*}{dA\; d\ln m}.
\eeq
It is customary in work on the IMF to express the results in terms of
base-10 logarithms, but we prefer natural logarithms since they are simpler to
use in theoretical calculations. For example,
Miller \& Scalo (1979; hereafter MS79) express the number of main-sequence
stars per unit
area between $m$ and $m+dm$ as $\phi_{\rm ms}(m) d\log m$; in our notation,
$\varsigma_{\rm ms}=\phi_{\rm ms}\log e=0.434\phi_{\rm ms}$.
We denote the volume density
per unit logarithmic mass interval
at the mid-plane by $n_{*0}(m)$. This is related to the
surface density by $\varsigma_*(m)=2H(m)n_{*0}(m)$, where $H(m)$ is the 
effective scale height of stars
of mass $m$. Since the scale height depends on the mass of the star,
the PDMF is not proportional to the volume density. 
The corresponding stellar mass
densities are $\Sigma_*(m)\equiv m\essm$ and $\rho_*(m)\equiv m n_*(m).$

	The  differential star formation rate (SFR) by number as a function of mass is
given by $d\edNsmt/d\ln m$, where $\edNsmt$ includes only
stellar births, not stellar deaths.
With our assumption that the IMF is universal, the 
differential SFR by number can be expressed as
the product of the total 
SFR by number, $\edNst(t)$, and
the probability that a star
is born with a mass $m$---i.e., the IMF, $\psi(m)$:
\beq
\frac{d\edNsmt}{d\ln m}=\edNst(t)\psi(m),
\label{eq:separable}
\eeq
where
\beq
\int_{m_\ell}^{m_u} \psi(m) d\ln m =1;
\eeq
recall that $m_u$ is the upper limit on the mass of an individual
star and $m_\ell$ is the lower limit. 
Note that the assumption of a universal IMF allowed us to write the star formation rate as
a separable function of mass and time, as did MS79 and Scalo (1986).

    The SFR by number per unit area of Galactic disk is then
\beq
\edss(m,t)=\frac{d^2\edNs(m,t)}{dA \, d\ln m}=\frac{d\edNst(t)}{dA}\psi(m)
	\equiv\edss(t)\psi(m)
\eeq
The ratio of the current SFR to the value averaged over the age
of the disk, $t_0$, is the parameter defined by MS79:
\beq
b(t_0)\equiv\frac{\edsst(t_0)}{\avg{\edsst}},
\eeq
where
\beq
\avg{\edsst}\equiv\frac{1}{t_0}\int_0^{t_0}\edsst(t)\;dt.
\eeq
Following MS79, who in turn followed Schmidt (1959), we note that
the PDMF is comprised of all the stars that are younger than their main-sequence
lifetime, $\tau(m)$, or the age of the Galaxy, and is related to the star formation rate by
\beq
\ess(m)=\int_{t_0-\tau'(m)}^{t_0}  \edss(t)\psi(m)\; dt
	=\avg{\edsst}\psi(m)\int_{t_0-\tau'(m)}^{t_0}  b(t)\; dt,
\eeq
where $\tau'(m) \equiv \min[\tau(m), t_0]$. Observe that by definition we have
\beq
\frac{1}{t_0}\int_0^{t_0} b(t)\; dt =1.
\eeq
In our notation, the number of stars ever born per logarithmic mass interval $d\ln m$  
(Schmidt 1959) is
\beq
\psi(m)\int\edsst(t)dt=\psi(m)\avg{\edsst} t_0=
0.434\xi(\log m),
\eeq
where $\xi(\log m)$ is the notation introduced by MS79.
Henceforth, for simplicity we 
shall generally omit the argument $t$ in $\edNsmt$ and
related quantities. 

\subsection{Previous Forms for the IMF}
\label{sec:forms}

Here we describe several forms for the IMF that have been proposed in the
literature 
(cf. Kroupa 2002, Bastian et al 2010).

\subsubsection{Salpeter IMF}

The original IMF was proposed by Salpeter (1955) and is the simplest:
\beq
\psi_{\rm Sal}\propto m^{-\Gamma}~~~~(m_\ell\leq m\leq m_u) 
\eeq
with $\Gamma\simeq 1.35$. The mean mass is 
\beq
\avg{m}=
\frac{\Gamma m_\ell}{\Gamma-1}\left[1-\left(\frac{m_\ell}{m_u}\right)^{\Gamma-1}\right]
	\simeq 3.5 m_\ell,
\label{eq:salpeter}
\eeq
where we have omitted a factor $(m_\ell/m_u)^\Gamma \ll 1$ and where the numerical evaluation is for
$\Gamma=1.35$.
The mass at which $\psi_{\rm Sal}$ reaches a peak (the mode of the IMF) is 
$\mpk=m_\ell$, which we fix to $m_\ell = 0.21$ to fit the ratio $R_{MK}$ (see \S 3).
This power law could be applied to either individual stars or to stellar systems.
The primary disadvantage of this form for the IMF is that it does  not
allow for the turnover in the IMF at low masses.

\subsubsection{Scalo IMF (effective)}

Scalo (1986) made a careful assessment of the data avaliable at that time and determined
the resulting IMF, which he presented in a table 
in the mass range $0.09-63$. 
The mean mass depends on the
age of the disk, and is in the range  $\avg{m}=0.41-0.52 $ for a disk age $t_0=9-12$~Gyr. 
The peak of the IMF is $\mpk=0.29$. 
For high-mass stars, this IMF was not corrected for the time these stars remain obscured by 
their parent clouds, nor for the necessity of including the largest OB associations, 
which are very rare
(Parravano et al. 2009), in order to obtain a fair sample of massive stars;
as a result, this IMF is expected to be deficient in high mass stars.
Scalo (1998a) proposed a three-power
law fit for the IMF in the range $0.1  \leq m\leq 100$, but we
shall not consider this form here.

\subsubsection{Kroupa IMF (individual/effective)}

Kroupa (2001, 2002) proposed a four-power law fit to the IMF. This form for the IMF appears to be the first in which $\psi$ has a power-law behavior at low masses with $\gamma>0$:
\beq
\psi_{\rm K}\propto m^{\gamma_i}~~~~\mbox{with}~~~~\left\{
\begin{array}{ll}
\gamma_0=\ \ 0.7\pm0.7~~~&(0.01\leq m <0.08)\\
\gamma_1=-0.3\pm0.5~~~&(0.08\leq m< 0.50)\\
\gamma_2=-1.3\pm 0.3~~~&(0.50\leq m<1.00)\\
\gamma_3=-1.3,-1.7 \pm 0.3~~~&(1.00\leq m).
\end{array}
\right.
\label{eq:kroupa}
\eeq
(Note that we have modified Kroupa's notation so that $\psi$ is in our units of  stars per
logarithmic mass interval.  Also note that in the limit of low mass, his $\gamma _0$ is our $\gamma$; at high
mass his $\gamma _3$ is our $-\Gamma$.)
Kroupa points out that this IMF is a hybrid, referring to individual stars at low masses and stellar
systems at higher mass ($m\ga 3 $). Based on the results
of Sagar \& Richtler (1991), he estimates that the individual-star IMF will be somewhat steeper at high masses, with $\Gamma\simeq 1.7$ as listed above (note, however, that
Ma{\'{\i}}z Apell{\'a}niz 2008 finds that binarity has a smaller effect on $\Gamma$).
The peak of this IMF occurs at $\mpk=0.08 $. 
In common with
the Chabrier IMF and the STPL IMF discussed below,
this form of the IMF relates the rate of formation of high-mass stars to that of
solar-type stars by a simple power law.

The advantage of this form for the IMF
is that it is an approximate analytic representation of the data. It is somewhat surprising that
the boundary between the lowest two mass ranges occurs at the brown-dwarf mass, however,
since there is no reason that hydrogen burning, which turns on long after the star forms, should
affect the mass of the star. Another difficulty with this form is that it is a hybrid, referring
to single stars at low masses and systems at high masses. 

\subsubsection{Chabrier IMFs (individual-star and effective)}

Chabrier (2005) presents both individual-star and system IMFs based primarily on the
observations of nearby stars by Reid et al. (2002). 
He finds
\beq
\psi_{\rm C}\propto \left\{ \begin{array}{ll}
\exp\left[-\frac{\dis(\log m - \log 0.20)^2}{\dis2\times (0.55)^2}\right] & ~~~~\mbox{(individual-star)}\\ \\
\exp\left[-\frac{\dis(\log m - \log 0.25)^2}{\dis2\times (0.55)^2}\right] & ~~~~\mbox{(effective)}
\end{array}
\right.
\label{eq:chabrier}
\eeq
for $m\leq 1 $ and
\beq
\psi_{\rm C}\propto m^{-1.35\pm 0.3}
\label{eq:chabrier-gt1}
\eeq
for $m>1 .$
These IMFs can be described as a ``log-normal/Salpeter" shape.
The peak masses and the mean masses of MS stars for these IMFs are 
$\mpk=0.20,\; \langle m \rangle_{\rm ms}=0.70$ and
$\mpk=0.25,\; \langle m \rangle_{\rm ms}=0.79$, respectively, for $m_u=120$ .

Earlier, Chabrier (2003b) proposed a log-normal/Salpeter IMF with
$\mpk=0.08$ for the individual-star IMF and $\mpk=0.22$ for the effective
system IMF.
As we shall see below, this form for the individual-star
IMF is not consistent with the data we have analyzed to infer the IMF.
Henceforth, we shall mean the Chabrier (2005) IMF whenever we
refer to the ``Chabrier IMF."

\subsubsection{Padoan-Nordlund IMF (theoretical system IMF)}

Padoan \& Nordlund (2002) and Padoan et al. (2007) proposed the first IMF based on theory
that approximately captures the observed behavior of the IMF at both low and high masses. 
The premises underlying this theory are
that the mass distribution of gravitationally unstable cores is generated by the process of turbulent fragmentation, and that the system IMF is determined by this core mass function.
The proposed shape of the IMF is (Padoan et al. 2007)
\beq
\psi_{\rm PN}(m) \propto  m^{-\Gamma} \left\{ 1 + {\rm erf} \left[\frac {4 \ln(m/\emch)+\sigma^2}{2 \sqrt{2} \sigma} \right] \right\},
\label{eq:PN}
\eeq
where $\mch$ is the Bonnor-Ebert mass in terms of the average density and temperature
and $\sigma^2$ is the variance of the gas density PDF.
Above 1 \msun, Padoan and Nordlund obtain a power-law distribution of dense cores consistent with the high-mass IMF
slope. At low masses, the mass distribution is determined by the PDF of the gas density, which is assumed to be lognormal. 
Padoan et al. (2007) show that for reasonable values of the physical parameters 
(turbulent Mach number $\calm_0 = 10$, density $n=10^4$ cm$^{-3}$, temperature $T=10$~K, and slope of the velocity
power spectrum $\beta=1.9$, which leads to 
$\Gamma=1.4$, $\emch=1$ and $\sigma=1.8$), the mass distribution in equation (\ref{eq:PN}) becomes similar to the Chabrier (2003b) effective system IMF,
which is similar to the Chabrier (2005) effective system IMF.
The resultant mode of the IMF and the mean mass of MS stars are respectively $\mpk=0.24$ and $\langle m \rangle_{\rm ms}=0.75$.
In reality, the PN theory addresses the distribution of core masses that in principle are larger than the stellar system masses they form. However, if the shape of the mass distribution of cores and stellar systems are similar, then equation (\ref{eq:PN}) with a larger $\emch$ gives the core mass distribution. Hennebelle and Chabrier (2008, 2009) assume that the core mass is 3 times the stellar system mass.

While this is an impressive attempt at a theory of the IMF, it has a number of
problems (McKee \& Ostriker 2007, Hennebelle \& Chabrier 2008, 2009). The basic idea
of their model is that fragments are generated by shocks that have a compression
that is linear in the shock velocity, as expected for shocks with velocities that
are large  compared to both the sound speed and the Alfven velocity; 
however
this assumption is valid only on large scales in large GMCs. They assume that
the number of fragments scales inversely as the cube of the size, but do
not justify this choice over several plausible alternatives.
Furthermore, their theory neglects
the turbulent structure in the cores that form massive stars (McKee \& Tan 2002, 2003;
Hennebelle \& Chabrier 2009).

\subsubsection{Hennebelle-Chabrier IMF (theoretical system IMF)}

Hennebelle \& Chabrier (2008, 2009) have developed a theory for the IMF based on
the Press-Schechter (1974) theory for cosmological structure formation.
It overcomes each of the problems cited above for the Padoan \& Nordlund theory.
However, in contrast to the Padoan-Nordlund theory, it ignores the magnetic field.
Like Padoan and Nordlund, they assume that that IMF is determined by the mass
function of the molecular cores out of which the stars form. Since they do
not follow fragmentation of the cores, their IMF is a system IMF, which they
take to be the same as the effective IMF. They assume that
the stellar system mass is 1/3 of the core mass based on observations suggesting that the core masses are several times greater than the resulting stellar masses (e.g., Alves, Lombardi
\& Lada 2007). For an isothermal equation of
state, their core mass function is (Hennebelle \& Chabrier 2008)
\beq
\psi_\hc\propto \frac{1}{(\mtil\Rtil^3)^{1/2}}\left[\frac{1+(1-\eta)\calm_*^2\Rtil^{2\eta}}
	{1+(2\eta +1)\calm_*^2\Rtil^{2\eta}}\right]\exp\left\{-\frac{[\ln(\mtil/\Rtil^3)]^2}
	{2\sigma^2}-\frac{\sigma^2}{8}\right\},
\eeq
where
\beq
\mtil\equiv \frac{m}{\emch}= \Rtil(1+\calm_*^2\Rtil^{2\eta}),
\eeq
$\emch$ is the Jeans mass,
$\Rtil$ is the radius of the clump in units of the Jeans length, $\sigma^2$ is
the variance of the gas density fluctuations, $\calm_*$ is the characteristic
Mach number at the Jeans scale, and $\eta$ is the
exponent of the linewidth-size relation. They take $\eta$ to be in the range
$0.4-0.45$ based on the simulations of Kritsuk et al (2007). At moderately high masses
($m\ga \emch$), $\mtil/\Rtil^3 \propto \mtil^{-0.6}$ to $\mtil^{-0.7}$ and
the mass dependence due to the $\exp\ln^2$ factor is weak; if that is ignored,
\beq
\psi_\hc\propto\mtil^{-(\eta+2)/(2\eta+1)}\equiv\mtil^{-\Gamma_\hc}~~~~~(\mtil\ga 1),
\eeq
which gives $\Gamma_\hc\simeq 1.3$ for the value of $\eta$ they adopt.  At very high
masses ($\mtil \gg 1$), the exponential term again takes over the mass dependence
and $\psi_\hc$ drops off more steeply with mass, reverting to a log-normal type distribution.
At low masses ($\mtil <1$), $\mtil/\Rtil^3 \propto \mtil^{-2}$ and the exponential term rises rapidly with increasing mass,
producing a log normal form again. 
Their results for a non-isothermal equation of
state are considerably more complicated, but yield results that are qualitatively consistent
with the isothermal theory (Hennebelle \& Chabrier 2009).

Both the Padoan-Nordlund IMF and the Hennebelle-Chabrier IMF apply to a particular
star-forming cloud. In order to determine an average IMF that can be compared
with observations of stars from different clouds, it is necessary to average the IMF
over a distribution of cloud temperatures, densities and Mach numbers. We 
assume that it is possible to represent this average by a suitable
choice of $\emch$,  $\calm_*$ and $\sigma$.
For the parameter values typically used by Hennebelle and Chabrier 
($\eta=0.4$, $\calm_*=\sqrt{2}$ and  $\sigma=1.9$), the mode of the distribution matches the Chabrier 
effective
value $\mpk=0.25$ for $\emch=1.56$. In this case, the mean mass of MS stars is $\langle m \rangle_{\rm ms}=0.88$.

\subsection{Smoothed Two-Power Law IMF}
\label{sec:stpl}

\begin{figure}
\epsscale{0.9}
\plotone{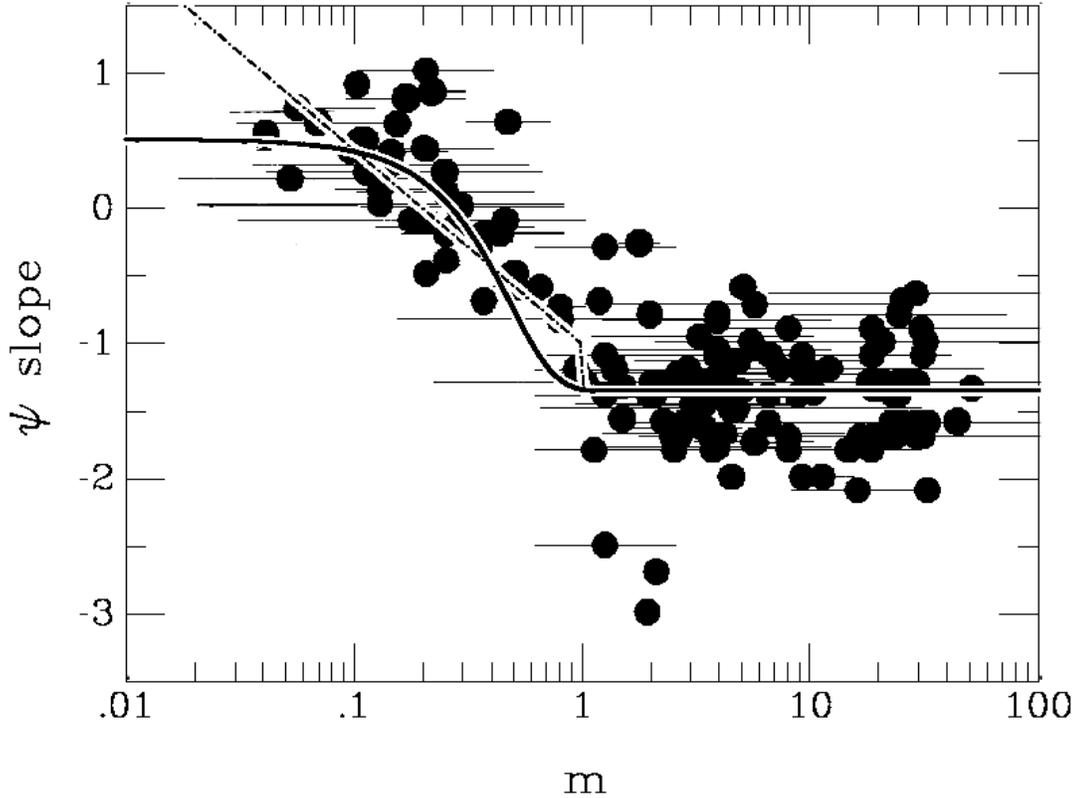}
\caption{Hillenbrand (2004) compilation of several studies providing
the IMF slope as function of the mass covered by the study.
The horizontal thin line through each data point is the 
range of masses over which 
the single slope is claimed to apply. 
The curves superimposed on the  Hillenbrand (2004) plot 
correspond to the mass dependence of the slope of two IMF functions:
The continuous curve is the $\psi_{\rm STPL}$ (eq. \ref{eq:psi}  with $\Gamma=1.35$, $\gamma=0.51$ and $\emch=0.35$), 
and
the dashed curve gives the slope of $\psi_{\rm C}$ (eqs. \ref{eq:chabrier}-\ref{eq:chabrier-gt1}).
}
\label{fig:alphaplot}
\end{figure}

Existing observations of the IMF are consistent with, but do not require,
power-law behavior at low masses as well as at high masses.
Figure 
\ref{fig:alphaplot}
shows the Hillenbrand (2004) 
compilation of several studies providing
the IMF slope as function of the mass covered by the study.
As she pointed out, 
the relatively small scatter 
at low masses is quite surprising given the variety of observational 
techniques and the variety of transformations between the observations 
and stellar masses that are used. Above masses of
several $M_\odot$, the scatter is larger, but the mean slope
does not appear to change
with mean mass and is centered on the Salpeter (1955) slope of -1.35. 
The Salpeter slope holds down to at least 1 \msun\ and believably down to 
0.6 \msun.

To capture a power-law behavior at low masses 
($\psi\propto m^\gamma$) and high masses ($\psi\propto m^{-\Gamma}$), we adopt
a simple analytic form, the Smoothed Two-Power Law (STPL) IMF, as discussed in \S 1.
This form for the IMF has a shape that is determined by three parameters plus a lower and upper mass cutoff:
\beq
\psi_{\rm STPL}(m)= C \, m^{-\Gamma} \left\{1-\exp\left[{-(m/\emch)^{\gamma +
\Gamma}}\right]\right\}
~~~~~(m_\ell\leq m \leq m_u).
\label{eq:psi}
\eeq
Each of the five parameters ($m_\ell,\;m_u,\; \emch,\; \gamma,$ and
$\Gamma$) has a direct physical significance;
the normalization factor $C$ is determined in terms of these parameters (see below).
We assume that star formation is suppressed below a
mass $m_\ell$ (possibly due to opacity effects in fragmentation---Low
\& Lynden-Bell 1976) and above a mass $m_u$.
The parameter $\emch$ 
approximately determines the position of the IMF maximum.
Finally, $\gamma$ and $-\Gamma$ are the low-mass and high-mass slopes
of the IMF, respectively. 
The advantages of this form for the IMF are that it
captures the observed power-law behavior at low and high masses and that
it has a simple analytic form. The disadvantage is that, unlike a log-normal,
for example, there is no parameter that independently determines the width
of the IMF; instead, the full-width half
maximum of the IMF is determined by the two power laws. Nonetheless, as
we shall see, it provides a good fit to the available data.

Since the STPL IMF has a simple, analytic form, it is possible to directly
calculate a number of quantities of interest:
The normalization factor $C$, which
ensures that $\int_{m_\ell}^{m_u}\psi_{\rm STPL}(m)\, d\ln m = 1$, is
\begin{eqnarray}
\frac{1}{C}&=&\int_{m_\ell}^{m_u}m^{-\Gamma} 
\left\{1-\exp\left[{-(m/\emch)^{\gamma + \Gamma}}\right]\right\}\, d\ln m,\\
&\simeq&
\frac{1}{\Gamma \emch^\Gamma}\left[ G\left(\frac{\gamma}{\gamma+\Gamma}\right)
-\frac{\Gamma}{\gamma}\left(\frac{m_\ell}{\emch}\right)^\gamma\right],
\end{eqnarray}
where $G(x)$ is the Gamma function and we have assumed that
$m_\ell \ll \emch\ll m_u$. 
The mean mass is
\beq
\avg{m}\simeq\frac{\Gamma \emch}{\Gamma-1}\left[\frac{\dis G\left(\frac{\gamma+1}
{\gamma+\Gamma}\right) - \left(\frac{m_u}{\emch}\right)^{-(\Gamma-1)}}
{ \dis G\left(\frac{\gamma}{\gamma+\Gamma}\right)-\frac{\Gamma}{\gamma}
\left(\frac{m_\ell}{\emch}\right)^\gamma}\right],
\eeq
while the peak mass (accurate to 4\% for $\Gamma\geq 1$ and $\Gamma/\gamma\geq 1.2$) is,
\beq
\mpk\simeq \emch \left[\sqrt{\frac{9}{4}+6\frac{\gamma}{\Gamma}}-\frac{3}{2}\right]^\frac{1}{\gamma+\Gamma}.
\eeq
The fraction of stars that are born as brown 
dwarfs (i.e., with masses less than $\embd$) is
\beq
F_\bd\simeq \Gamma\left[\frac{\dis \left(\frac{\embd}{\emch}\right)^\gamma
	-\left(\frac{m_\ell}{\emch}\right)^\gamma}
	{ \dis \gamma G\left(\frac{\gamma}{\gamma+\Gamma}\right)-\Gamma
\left(\frac{m_\ell}{\emch}\right)^\gamma}\right],
\label{eq:fbd}
\eeq
which is accurate to better than 2\%,
and the mass fraction of stars born as brown dwarfs is
\begin{eqnarray}
F_{m,\,\bd}&=&\frac{1}{\avg{m}}\int_{m_\ell}^{m_\bd} m\psi\; d\ln m,\\
&\simeq& \frac{\Gamma-1}{\gamma+1}\left[\frac{\dis \left(\frac{\embd}{\emch}\right)^{\gamma+1}
	-\left(\frac{m_\ell}{\emch}\right)^{\gamma+1}}
	{ \dis G\left(\frac{\gamma+1}{\gamma+\Gamma}\right)-\left(\frac{m_u}{\emch}\right)
	^{-(\Gamma-1)}}\right],
\end{eqnarray}
which is accurate to better than 5\%.

\subsubsection{De Marchi-Paresce-Portegies Zwart IMF (effective IMF)}

As discussed in the Introduction, Paresce \& De Marchi (2000), 
De Marchi \& Paresce (2001), and De Marchi, 
Paresce \& Portegies Zwart (2003, 2010, hereafter MPPZ) have applied the
STPL form of the IMF to both young galactic clusters and globular clusters.
They make no attempt to correct for binarity, so theirs is an effective system IMF.
They found that the characteristic mass
of a star in
a cluster increases with time due to dynamical
evolution.
Over a mass range $0.1<m<10$,
De Marchi et al (2010) infer $\Gamma=0.97\pm 0.17$, $\gamma = 1.54\pm 0.64$,
and, at birth, $\mch=0.15$.

\begin{figure}
\epsscale{0.9}
\plottwo{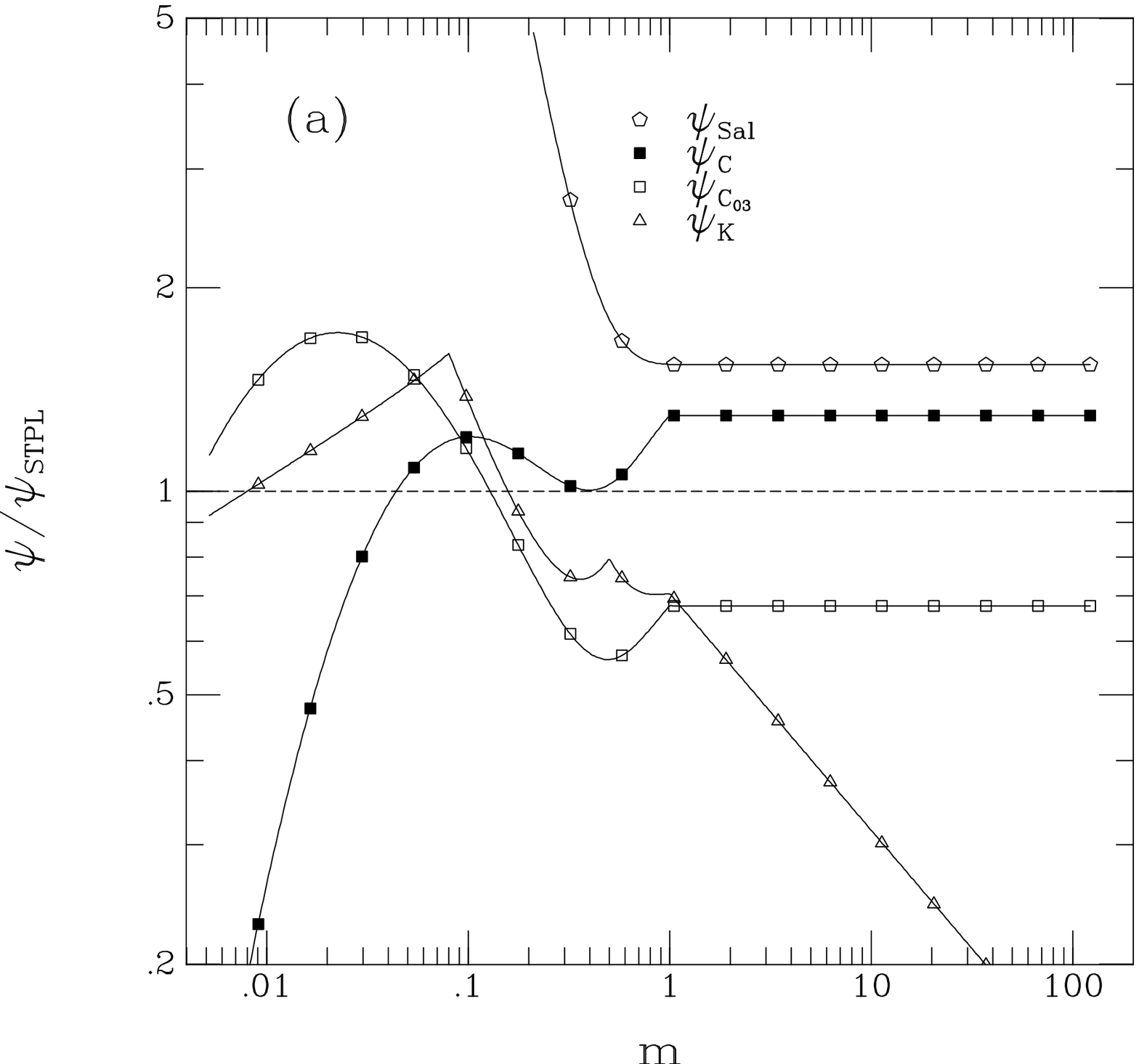}{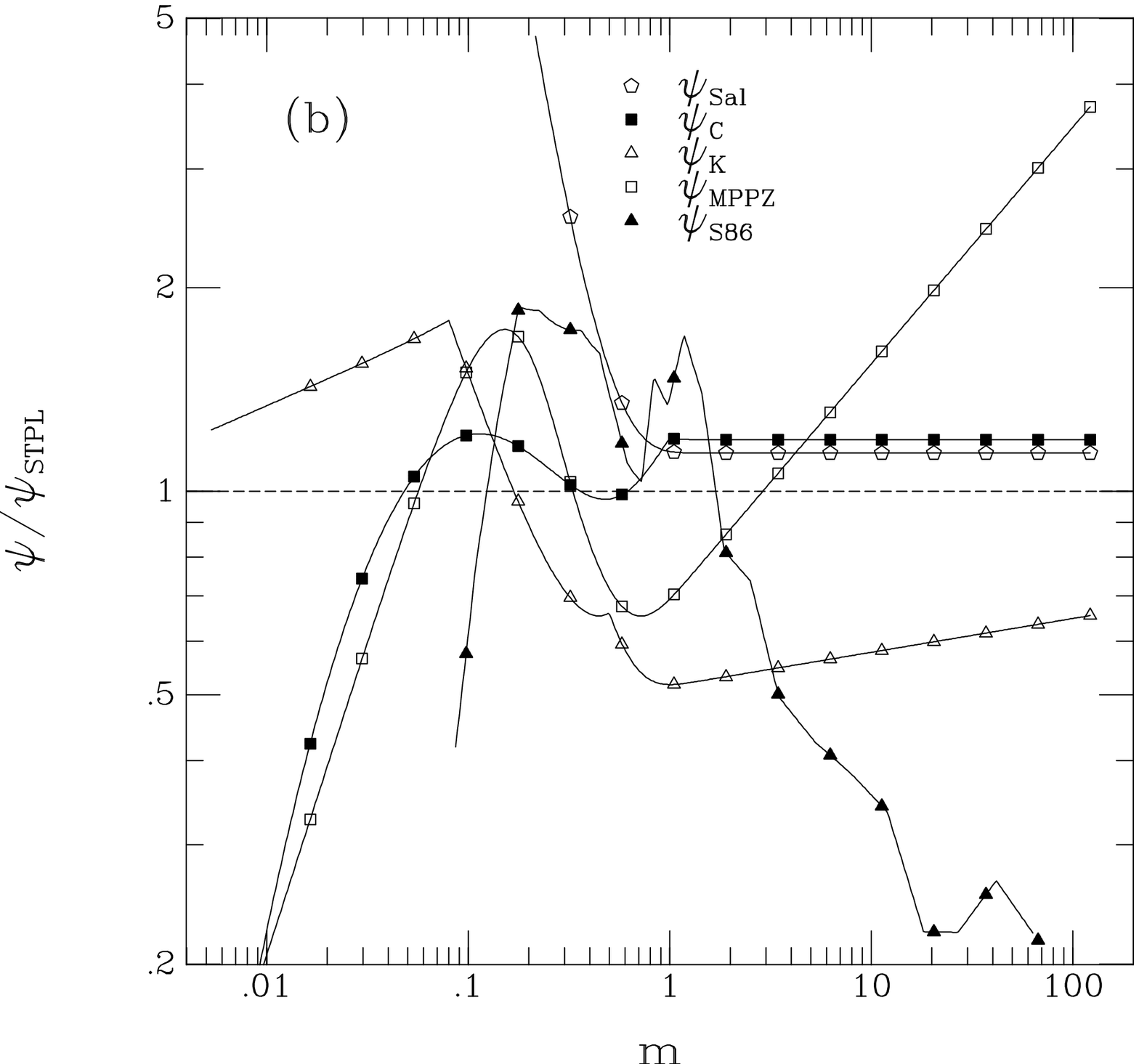}
\plotone{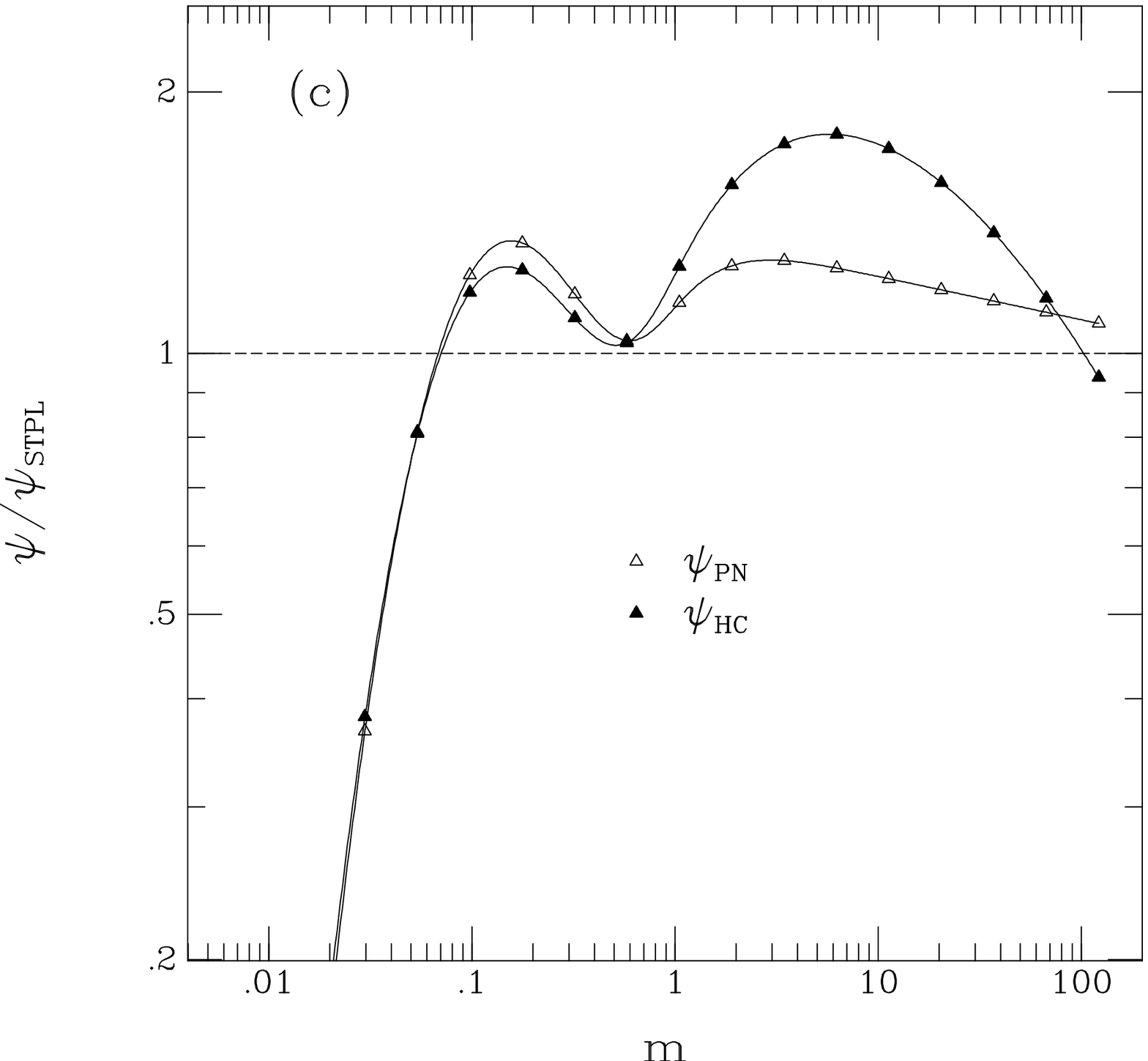}
\caption{
Different forms of the IMF normalized to $\psi_{\rm STPL}$. 
a) Individual-star IMFs $\psi_{\rm Sal}$, $\psi_{\rm C}$,  $\psi_{\rm C_{03}}$ and $\psi_{\rm K}$ normalized to $\psi_{\rm STPL}(\Gamma=1.35,\gamma=0.51,\emch=0.35)$. $\psi_{\rm C_{03}}$ refers to the individual-star IMF proposed by Chabrier (2003a, 2003b), which has the same form than $\psi_{\rm C}$ but with $m_{\rm peak}=0.08$ and $\sigma=0.69$. 
b) Effective system IMFs $\psi_{\rm Sal}$, $\psi_{\rm C}$, $\psi_{\rm K}$, 
$\psi_{\rm MPPZ}$ and $\psi_{\rm S86}$ normalized to 
$\psi_{\rm STPL}(\Gamma=1.35,\gamma=0.57,\emch=0.42)$. $\psi_{\rm S86}$ is the IMF in Scalo (1986)
Table 7 for a disk age of 12 Gyr and $b_0=1$.
c) Theoretical system IMFs $\psi_{\rm PN}$ and  $\psi_{\rm HC}$  normalized to 
$\psi_{\rm STPL}(\Gamma=1.35,\gamma=0.57,\emch=0.42)$. 
For all IMFs  $\int \psi d\ln m = 1$ in the range $0.004<m<120$, except for $\psi_{\rm Sal}$ and $\psi_{\rm S86}$ for which the mass ranges are respectively $0.21<m<120$ and $0.09<m<63$.
}
\label{fig:ratio-imfs}
\end{figure}

The shape of this IMF and the others considered in this paper are shown in 
Figure \ref{fig:ratio-imfs}
for the parameter values given in Table 2.

\newpage 

\section{Observational Determination of IMF Parameters (Individual-Star)}
\label{sec:obsdet}

\subsection{The High-Mass Slope $\Gamma$}
\label{sec:Gamma}

The STPL form for the IMF has a single value of $\Gamma$ for
intermediate and high-mass stars.
The value of $\Gamma$ for clusters and associations shows considerably 
scatter but the average value of $\Gamma$ is close to the Salpeter value.
For clusters, Scalo (1998a) concluded that $\Gamma$ for 
1-10 $M_\odot$ is 1.5 to 1.8, but probably flatter ($\sim 1.2$)
at larger masses.
A fit to the high-mass end of Scalo's (1986) IMF gives $1.5<\Gamma<1.7$.
However, one obtains $1.3<\Gamma<1.5$ when Scalo's (1986) IMF 
is corrected by 
considering that (i) short-lived
stars spend part of their lifetime obscured by 
their parent clouds and (ii) in order to obtain a fair sample of massive stars,
it is necessary to include the largest OB associations, which are very rare 
and therefore often omitted in Galactic surveys (Parravano et al. 2003, 2009).
Additionally, Elmegreen and Scalo (2006) pointed out that if the SFR were decreasing 
because we are moving farther from the density wave shock, but one assumed that it was 
constant in determining the IMF from the PDMF, then one would get a value that is too steep. 
This effect adds to the above two corrections in the same direction.

When the various estimates
of the IMF slope derived from star counts of clusters and associations 
are plotted as function of the average
mass over which a common slope is claimed to apply
(Scalo 1998a, Kroupa 2001, Hillenbrand 2004; see Fig. 1), the data 
show considerable scatter for $m > 1$,
but the average value of $\Gamma$ is close to the Salpeter value of 1.35. 
Elmegreen (2006) reviewed observations of galaxy-wide IMFs and concluded that,
with rare exceptions, the high-mass slope averaged over
entire galaxies is consistent with that observed
in star clusters, which is about the Salpeter value. More recently, 
Bastian et al (2010) concluded that the available data are indeed consistent with $\Gamma\simeq 1.35$.
De Marchi et al. (2010) find $\Gamma\simeq 1$ for young clusters in their sample, but
it must be borne in mind that they are focusing on stars with $m<10$.

How does binarity affect the high-mass slope?
Ma{\'{\i}}z Apell{\'a}niz (2008) has summarized the evidence
that most massive stars are in binaries. However,
through numerical experiments, he shows that this does not have
a significant effect on the high-mass slope of the IMF.
For our standard model we therefore adopt $\Gamma=1.35$, but we explore
the consequences of changing $\Gamma$ to 1.2 and 1.5.

\subsection{Upper Mass Limit $m_u$}
\label{sec:mu}

	The existence of an upper limit to the mass function
has been controversial; for example, Massey \& Hunter (1998) suggested
that the apparent cutoff is due to the scarcity of very massive
stars.  This issue was addressed in a general way by MW97 and Parravano
et al. (2003), who showed that the cutoff is physically significant 
for various samples of stars in the Galaxy.  
Elmegreen (2000) argued that the data support the existence of
an upper mass limit and proposed a model that shows 
a high-mass turndown in the IMF.  Oey \& Clarke (2005) showed that observations
of a number of star clusters indicate that $m_u<200$.
In a detailed study of the Arches Cluster, Figer (2005) concluded that
$m_u<150$. These studies do not take multiplicity into account;
for example, a star that was thought to violate this upper mass limit,
Pismis 24-1, with an inferred mass exceeding $200 M_\odot$, has been resolved into
three stars with masses in the range $92-97 M_\odot$ (Ma{\'{\i}}z Apell{\'a}niz
et al. 2007). The highest dynamically determined stellar masses are those in
a Wolf-Rayet binary, with masses of $83\pm 5 M_\odot$ and $82\pm 5 M_\odot$
(Bonanos et al. 2004); note that these are present-day masses, not initial ones.
Ma{\'{\i}}z Apell{\'a}niz (2008) summarizes the evidence that the upper mass limit
is $m_u\simeq 120$.

For our standard model we therefore adopt 
$m_u=120$ for {\it individual stars}, but we explore the consequences of adopting values 
from $m_u=80 - 150$.\footnote{A very recent paper by Crowther et al. (2010), utilizing observations
of the young star clusters NGC3603 and R136, suggests that $m_u$ may be as large as 300.
As we will show, large values of $m_u$ have very little effect on the conclusions of this paper, and we
maintain our range 80$< m_u < 150$ here.} In reality, the decrease in the IMF near the upper mass
limit is likely to be smooth rather than abrupt, so that our model of an IMF
with a sharp cutoff at $m_u$ is approximate. The global effects of massive stars
are often inferred from their ionizing luminosity, but the existence of a few stars
with masses exceeding $m_u$ 
in a cluster or galaxy with a very large number of massive stars
does not have a significant effect on
the total ionizing luminosity.
As stated above, massive star formation might be suppressed in the
outer parts of galaxies, but most of the star formation occurs in the disk regions where massive stars also form.

\subsection{Lower Mass Limit $m_\ell$}

Theoretically, opacity effects should limit fragmentation into objects smaller
than some critical value (Low \& Lynden-Bell 1976); including the effects of He,
this is about $0.004 M_\odot$ (Whitworth et al. 2007). Caballero et al. (2007)
have found that the mass distribution of brown dwarfs in one cluster extends smoothly
down to $0.006 M_\odot$, the lowest limit observed to date, which
is consistent with this theoretical limit.  We adopt $m_\ell= 0.004$.
Because the IMF is rising with mass at very low masses, the exact value
of this lower limit does not appreciably affect our estimates of the number
or mass fraction of brown dwarfs.

\subsection{The Shape of the Low-Mass Main-Sequence IMF}
\label{sec:lowmassshape}

	To determine the characteristic mass, $\emch$, and the low-mass
power law, $\gamma$, we first consider the shape of the low-mass,
main sequence part of the IMF. We focus on stars
that have sufficiently low masses 
($m< 0.8 $)
that their lifetimes exceed the age of the disk; 
thus, the IMF is the same as the PDMF for these stars.
We use one parameter to describe this part of the observed IMF,
\beq
R_{MK}=\frac{\caln_*(m=0.1-0.6)}{\caln_*(m=0.6-0.8)},
\label{eq:rmk}
\eeq
which is very roughly the ratio of the number of M stars to K stars.

The low-mass, main-sequence IMF can be inferred using local field 
stars, distant field stars, or cluster stars.
Each data source has its advantages.
We shall use field stars in the local neighborhood, which have
the advantages that many of the binaries are resolved and the distances are directly
measured, so that the absolute magnitude is known. The dominant uncertainty
in this case is the relation between the mass of a star and its absolute magnitude
in whatever photometric band used for the survey. 
Samples of distant field stars can be far larger (e.g., Zheng et al. 2001; Covey et al. 2008;  
Bochanski et al. 2010), but have 
additional disadvantages: First, a metallicity-dependent 
color-magnitude relation must be used to estimate absolute magnitudes and distances, 
and second,
one must 
statistically correct for binaries to estimate the individual-star mass function. 
Clusters have the
advantage that all the stars are at the same distance. However, they may have systematic
deviations from the average IMF due to the circumstances of their formation and
again, a correction for binaries must be made (nonetheless, we shall
use observations of clusters to estimate the contribution of brown
dwarfs to the IMF--see \S \ref{sec:bd} below). Local Galactic clusters are generally
not large and so are subject to statistical uncertainties;
globular clusters provide large samples
of stars, but are subject to evolutionary effects (e.g., De Marchi et al. 2010).

We use the local sample of field stars created by
Reid et al. (2002, hereafter RGH02) based on
data from the Hipparcos catalog and the Palomar/Michigan State University  (PMSU)
survey. The PMSU volume-complete sample ($16\geq M_V \geq 8$) includes 558 main-sequence stars 
in 448 systems with distance limits ranging from 5 pc ($M_V=15.5$)
to 22 pc ($M_V=8.5$).
The Hipparcos 25 pc sample ($M_V \leq 8$) includes 538 single stars, 204 binaries, 
22 triples and four quadruple systems, corresponding to 1028 stars in 768 systems.
The luminosity function created from these data includes a correction for undetected
binaries in the Hipparcos sample: RGH02 find a multiplicity of only
30\% in that sample, about half that Duquennoy \& Mayor (1991) inferred for 
the multiplicity of solar-type dwarf stars, and they therefore doubled the contribution
of the measured secondaries. This correction has a maximum effect of 18\% in
a single magnitude bin, and the typical effect is less than 12\%.

We opted not to use the recent very large sample of distant field stars by Bochanski et al. (2010) 
for the reasons just cited as well as the following.
They analyzed about 15 million low-mass stars (0.1 $< m <$ 0.8) from the Sloan Digital Sky Survey and estimated the mass function of field low-mass dwarfs.
However, we note that close to the upper mass limit, the adopted mass-$M_J$ relation (Delfosse et al. 2000; hereafter D00) needs to be extrapolated for $M_J<5.5$ (i.e. $m_{\rm eff}>0.72$). 
In addition, the binary correction used to estimate the single-star from the system mass function does not account for the secondary stars in systems with effective mass above 0.8 \msun. For these reasons the individual-star mass function estimated by Bochanski et al. (2010) is likely to be an underestimate above $0.7 \emsun$.
This is probably the reason why the lognormal fit to this mass function has about the same peak mass, $m_{\rm peak}\sim 0.18 \pm 0.02$,  as Chabrier (2005, see eq. \ref{eq:chabrier}) but a smaller dispersion, $\sigma=0.34 \pm 0.05$ instead of 0.55. Compared with the Pleiades mass function (Moraux et al. 2004), the two mass bins above $m \sim 0.6$ in the Bochanski distribution 
(see Fig. 27 in Bochanski et al. 2010) 
are about a factor 0.5 below the Pleiades distribution.
Therefore, it is not possible infer an accurate value for $R_\MK$ from the results of Bochanski et al. (2010).
In \S 5.1 we compare the 
effective
mass function of distant field stars to our estimation of the effective IMF and discuss some possible sources of discrepancy. However, we emphasize again here that the main goal 
in this paper is to constrain the individual-star IMF.

To determine $R_{MK}$ and its uncertainty,
we use the luminosity function found by RGH02 for a local sample of field stars 
together with the two mass-luminosity relations they used to derive the mass distribution. 
One of the $m(M_V)$ relations is based on Delfosse et al. (2000) for $M_V>10$ and on their own empirical $m(M_V)$ relation for $M_V<10$ [hereafter denoted as the RGH $m(M_V)$ relation]; for this relation, masses 0.1, 0.6 and 0.8 \msun correspond to $M_V= 16.4$ , 8.84 and 7.0 magnitudes, respectively. The other mass-luminosity relation used by RGH02 is the Kroupa, Tout, \& Gilmore (1993) semi-empirical mass-luminosity relation [hereafter, the KTG $m(M_V)$ relation]; for this relation, masses 0.1, 0.6 and 0.8 \msun correspond to $M_V= 16.13$, 8.7 and 6.24 magnitudes, respectively. These two $m(M_V)$ relations give about the same number of stars in the mass range 0.1-0.6 \msun, but  the KTG relation includes more stars than the RGH
one in the mass range 0.6-0.8 \msun. With the KTG $m(M_V)$ relation, the RGH02 luminosity function implies $R_{MK}\simeq 8$, whereas for the RGH $m(M_V)$ relation, $R_{MK}\simeq 11$. We assume that $R_{MK}$ is between these two values, so that
\beq
R_{MK}=9.5\pm 1.5
\label{eq:rmk1}
\eeq
for the IMF of individual stars.

\subsection{Ratio of Low-Mass Main-Sequence Stars to Brown Dwarfs}
\label{sec:bd}

Uncertainties in the IMF grow as stellar masses decrease
below the hydrogen-burning mass limit ($\sim 0.08$ \msun) into 
the realm of brown dwarfs. 
The derivation of the mass function of field brown dwarfs from a survey in a given band involves the use of theoretical cooling curves parameterized by mass and age (Burrows et al. 2001; Baraffe et al. 2003), and a star formation history (Chabrier 2002).
Furthermore,  due to the intrinsic low luminosity of evolved brown dwarfs, the numbers of objects in these surveys are relatively small.
On the other hand, 
very young embedded clusters are particularly valuable because young brown dwarfs have relatively high luminosities 
and can be readily detected using infrared photometry. However, mass functions derived by placing the young, low-mass stars on the theoretical 
H-R diagram using spectroscopic and photometric observations are sensitive to the theoretical pre-main sequence (PMS) model used to infer the masses and ages of the stars. Furthermore, it is often assumed that all the stars in a cluster have the same age, whereas it is possible that the star formation extends over several dynamical times (Tan, Krumholz, \& McKee 2006).
Despite these uncertainties, nearby young clusters provide the best opportunity to 
determine the IMF in the low-mass and brown-dwarf regime and test whether it might be universal.

We note that some theories suggest that the low-mass IMF might not be universal:
In the Padoan \& Nordlund (2004) theory of brown-dwarf formation, for example,
brown dwarfs form in rare, high-density peaks, and therefore 
their abundance could be exponentially sensitive
to the ambient conditions  such as the Mach number of the supersonic
turbulence (McKee \& Ostriker 2007).
It is thus of great interest to determine the IMF at very low masses.

Observations have established that the IMF peaks at subsolar masses and
declines into the brown-dwarf regime. For example,
Luhman et al. (2000) compared the IMFs derived from the
H-R diagram for the Trapezium cluster by using two different PMS models. 
With the D'Antona \& Mazzitelli (1997) PMS model,  the IMF has a turnover at 
$\sim 0.3$ \msun, whereas with the Baraffe et al. (1998) model,
the turnover occurs at $\sim 0.6$ \msun.
Muench, Lada, \& Lada (2000) and Muench et al. (2002) derived the Trapezium
IMF from the infrared (K-band) luminosity function and found a
turnover  around 0.15 \msun. 
Despite the quantitative differences between these IMFs,  they all show that
the mass distribution $d\caln_*/d\ln m$
has a broad maximum at subsolar
masses, with a turnover between 0.1 and 0.6 \msun, and then 
declines with decreasing mass in the
brown-dwarf range.
Bastian et al (2010) conclude that the existing data show that the 
IMF peaks in the mass range $(0.2-0.3) M_\odot$ and then declines with $\gamma> 0.5$.

The fraction of stars that are brown dwarfs sets a strong constraint on the IMF that 
is relatively insensitive to the effects of uncertainties in the PMS tracks.
This fraction can be determined from observations of
nearby young star clusters, which provide samples that extend well below the
hydrogen-burning limit. 
Since these data do not take binarity into account, they provide a constraint on
the effective IMF.
We follow Andersen et al. (2006, 2008) in characterizing 
the low-mass IMF function by the ratio 
of the number of subsolar main-sequence stars
to that of readily observable ($m>0.03$)
brown dwarfs in nearby clusters,
\beq
R_{\rm bd}^\eff=\frac{\caln_*(m=0.08-1)}{\caln_*(m=0.03-0.08)}.
\label{eq:rbd}
\eeq
Andersen et al. (2006, 2008) approximated the transition from the main sequence to brown dwarfs as occurring at $m_\bd=0.08$. We adopt this value for evaluating $R_\bd$,
but for all other properties of the population of brown dwarfs, we use the
best theoretical value, $m_\bd=0.075$ (Burrows et al. 2001).
Note that this ratio is  not influenced by possible variations of the very low 
cutoff mass of the IMF 
because $m=0.03$ is significantly above the expected values of this cutoff ($\sim 0.004 -0.006 \emsun$; Whitworth et al. 2007, Caballero et al. 2007).

\subsubsection{Universality of the Brown-Dwarf IMF}
\label{sec:univ}

In order to use the ratio $R_\bd^\eff$ to determine the IMF, it must have a universal value.
Andersen et al. (2006) concluded that the brown-dwarf ratios they analyzed,
including those from field-star PDMFs, are
all consistent with a single value to within the errors. 
Andersen et al. (2008) showed that the values of
$R_{\rm bd}^\eff$ for seven clusters (Taurus, ONC, Mon R2, Chamaeleon, Pleiades, NGC 2024 and IC 348) are statistically consistent with
the Chabrier (2005) lognormal IMF, a result that is consistent with
a universal IMF in the brown-dwarf regime. Here we strengthen this conclusion
by adding more 
cluster data and by focusing on the universality of 
$R_{\bd}^\eff$, not on a particular IMF shape.

In addition to the seven clusters considered by Andersen et al. (2008), we include
data from recent studies of NGC 1333 (Scholz et al. 2009) and NGC 6611 (Oliveira et al. 2009). 
Table 1 gives the brown-dwarf ratios, $R_{i,{\rm obs}}^\eff$, and the number of objects,
$\caln_{i,{\rm obs}}$ ($0.03<m<1$), observed in each cluster, labeled ``$i$".
Note that $\caln_{i,{\rm obs}}$ is the number of detected objects in
the surveyed area, not the actual number of objects in cluster $i$.
The estimated errors in $R_{i,{\rm obs}}^\eff$ agree with the expected root $\caln$ statistical fluctuations in the number of brown dwarfs ($\caln_{\bd}$) in the mass range $0.03<m<0.08$ and in the number of subsolar stars in the mass range $0.08<m<1$.

\begin{deluxetable}{llrl}
\tablecolumns{4}
\tablewidth{0pc}
\tablecaption{Ratio of Subsolar Stars to
Brown Dwarfs in Young Clusters}
\tablehead{
\colhead{Cluster} & \colhead{$i$} & \colhead{$\caln_{{i,\rm obs}}$} & 
\colhead{$R_{i,{\rm obs}}^\eff$} 
}
\startdata
 ONC & 1 & 185 & $3.3^{+0.8}_{-0.7}$ \\
 Pleiades & 2 & 200 & $4.9^{+1.5}_{-1.2}$\\
 NGC 2024 & 3 & 50 & $3.8^{+2.1}_{-1.5}$\\ 
 Taurus & 4 & 112 & $6.0^{+2.6}_{-2.0}$\\
 Chamaeleon & 5 & 24 & $4.0^{+3.7}_{-2.1}$\\
 IC 348 & 6 & 168 & $8.3^{+3.3}_{-2.6}$\\
 Mon R2 & 7 & 19 & $8.5^{+13.6}_{-5.8}$\\
 NGC 1333 & 8 & 68 & $2.7^{+1.0}_{-0.6}$\\
 NGC 6611 & 9 & 160 & $4.4^{+0.9}_{-0.6}$\\
\enddata
\end{deluxetable}
\clearpage

The value of the brown-dwarf ratio corresponding to the total observed sample is
\beq
R_{\bd,{\rm tot}}^\eff=\frac
{\sum_i \caln_{i,{\rm obs}} R_{i,{\rm obs}}^\eff/(1+R_{i,{\rm obs}}^\eff)}
{\sum_i \caln_{i,{\rm obs}}/(1+R_{i,{\rm obs}}^\eff)}=4.6^{+0.55}_{-0.47},
\label{eq:rbdlump}
\eeq
where a root $\caln$ statistical error was assumed in the numerator and the denominator of  equation (\ref{eq:rbd}).
We note that this result is consistent with the Chabrier (2005) effective IMF, which
has $R_{\bd}^\eff=5$.

To determine if the nine values of $R_{i,{\rm obs}}^\eff$ in Table 1 are consistent with a single universal value, $R_{\rm bd,univ}^\eff$, and, in particular, if they are consistent with the single value $R_{\rm bd,tot}^\eff$,
we compare the observed sample $\lbrace R_{\rm bd}^\eff\rbrace_{\rm obs}$ in Table 1 to 1000 synthetic samples $\lbrace R_{\rm bd}^\eff\rbrace_{\rm syn}$, each one consisting of nine synthetic clusters that are generated by imposing statistical fluctuations on the observed ones. Let $R_{i,k}$ be the value of $R_{\rm bd}$  for cluster $i$ in the
synthetic sample $\lbrace R_{\rm bd}^\eff\rbrace_{k}$, 
where $1 \leq i \leq 9$ and $1 \leq k \leq 1000$. The values of $R_{i,k}$ are
generated by Poisson fluctuations on the number of subsolar stars and on the number of
brown dwarfs observed in cluster $i$.
That is,
\beq
R_{i,k}=\frac{{\cal F}[\caln_{i,{\rm obs}} \, \, x/(1+x)]}
{{\cal F}[\caln_{i,{\rm obs}}/(1+x)]},
\label{eq:rbdki}
\eeq
where $x$ is the assumed universal value of $R_{\rm bd}^\eff$, and ${\cal F}(n)$ is an integer number
drawn from a Poisson distribution of mean equal to $n$.

The value  $R_{\rm bd,univ}^\eff$, or $x$, is
varied from 3 to 6.7 in steps of 0.0025. For each value of $x$ the following steps are performed:
(1) 1000 synthetic samples $\lbrace R_{\rm bd}^\eff\rbrace_{k}$ are generated, where each synthetic sample $k$ consists of nine $R_{\rm bd}^\eff$ values generated with equation 
(\ref{eq:rbdki}).
(2) The significance level $P(\{R_{\rm bd}^\eff\})$ 
of the Kolmogorov-Smirnov test is calculated for the
observed sample $\lbrace R_{\rm bd}^\eff\rbrace_{\rm obs}$ in Table 1 and for each of the 1000 synthetic samples, $\lbrace R_{\rm bd}^\eff\rbrace_{k}$. A cumulative distribution of a Gaussian centered at $x$ and with standard deviation 
$\sigma=0.3$ 
is used in the Kolmogorov-Smirnov test.
(3) The values of $P(\lbrace R_{\rm bd}^\eff\rbrace_{k})$ for the 1000 synthetic samples are ranked in descending order 
(4) The position of the 
significance level
for the observed sample $P(\lbrace R_{\rm bd}^\eff\rbrace_{\rm obs})$ in the above ranking list is obtained.

\begin{figure}
\epsscale{0.9}
\plotone{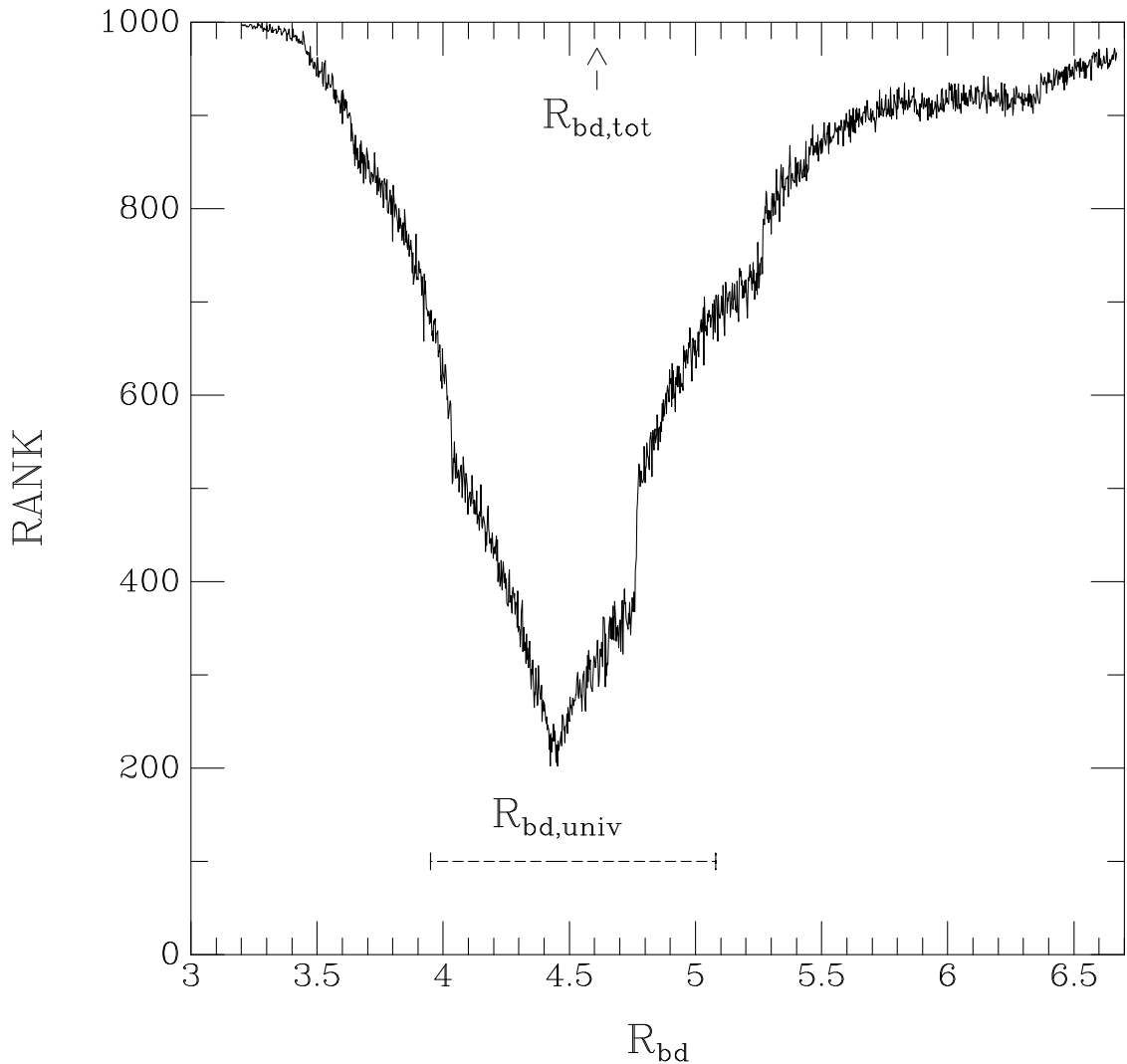}
\caption{Position of the observed sample in the list of the 1000 synthetic samples when arranged in decreasing order of KS-test significance levels.
The numerical experiment was repeated for $R$ at 0.0025 intervals. For the KS-tests a Gaussian cumulative distribution with $\sigma=0.3$ was used.
}
\label{fig:1}
\end{figure}

Figure \ref{fig:1} shows how the observed sample ranks among the Kolmogorov-Smirnov tests of the 1000 synthetic samples (the rank is 1 for the best significance level)
for each value of $x$. 
The noise-like fluctuations reflect the random changes in the 1000 synthetic samples.
The asymmetry of rank $(R_{\rm bd,univ}^\eff)$ is due to the fact that larger Poisson fluctuations occur in the number of brown dwarfs (for $R_{\rm bd}^\eff>1$), and  these fluctuations increase with increasing $R_{\rm bd,univ}^\eff$. In any case, there is a pronounced minimum close to $R_{\rm bd,tot}^\eff$, suggesting that $R_{\rm bd,univ}^\eff\simeq R_{\rm bd,tot}^\eff$ is highly probable, especially because the rank is in the top $20\%$. This method does not allow us to quantify the error in the value of $R_{\rm bd,univ}^\eff$. 
However, as shown in Figure \ref{fig:1}, we can quote  for example the range of $R_{\rm bd,univ}^\eff$ for which the observed sample ranks in the top 66\%; that is, $3.95 < R_{\rm bd,univ}^\eff < 5.1$.
Our results are fully consistent with those of Andersen et al. (2008): They
found that 63\% of their simulations had a probability higher than what is found for a Chabrier IMF, which has $R_\bd^\eff=5$, while the
results in Figure \ref{fig:1} show that the rank of the observed sample for $R_{\rm bd}^\eff= 5$ is $\sim 650/1000$.
We conclude that the mass distributions in the observed sample are consistent
with a universal value of the brown-dwarf fraction $R_{bd}^\eff$.

\subsubsection{The Brown-Dwarf Ratio $R_{\bd}^\ind$ for the Individual-Star IMF}
\label{sec:individual}

To infer the individual-star IMF from the constraints $R_{\MK}$, $R_{\bd}^\eff$ and $\Gamma$, it is necessary to correct the ratio $R_{\bd}$ for unresolved binaries. 
The ratio $R_{\MK}^{\ind}=9.5 \pm 1.5 $ obtained in \S \ref{sec:lowmassshape}
was estimated from the mass function of nearby individual stars, whereas the ratio $R_{\rm bd,univ}^\eff=4.6^{+0.5}_{-0.65}$ estimated in \S  \ref{sec:bd} 
was obtained from observations of young clusters, in which most binary systems are unresolved. 
In Appendix A we make a series of assumptions to account for binaries and then allow for the difference between the effective and the actual masses of stars to find (see eq. \ref{eq:rbdhind}) that 
$R_\bd^\ind \simeq 1.4 R_\bd^\eff/(1.2+\frac 19 R_\bd^\eff)$.  
Therefore,  the effective brown-dwarf ratio $3.95 < R_\bd^{\eff} < 5.1$ corresponds to $3.37 < R_\bd^{\ind} < 4.04$, so that
\beq
R_\bd^{\ind}= 3.76^{+0.28}_{-0.39} .
\label{eq:rbdh1}
\eeq
(Note that in obtaining the range of $R_\bd^{\ind}$, we ignored the errors in the parameters in equation \ref{eq:rbdhind}.)
Although Chabrier's (2005) fit to the IMF focused on stars with $m>0.1$,
our estimate for $R_{\rm bd,univ}^{\eff}$ agrees
with the value for the Chabrier 
effective IMF, $R_{\rm bd}^\eff\simeq 5$,
and our estimate for $R_{\rm bd,univ}^{\rm ind}$ agrees
with the value for the Chabrier individual-star IMF, $R^\ind_{\rm bd}\simeq 4$.
Chabrier's values for $R_\bd$ lie near the upper end of the values we infer from the data.

\subsection{Results: Observational Determination of $\mch$ and $\gamma$ }

\begin{figure}
\epsscale{1.0}
\plotone{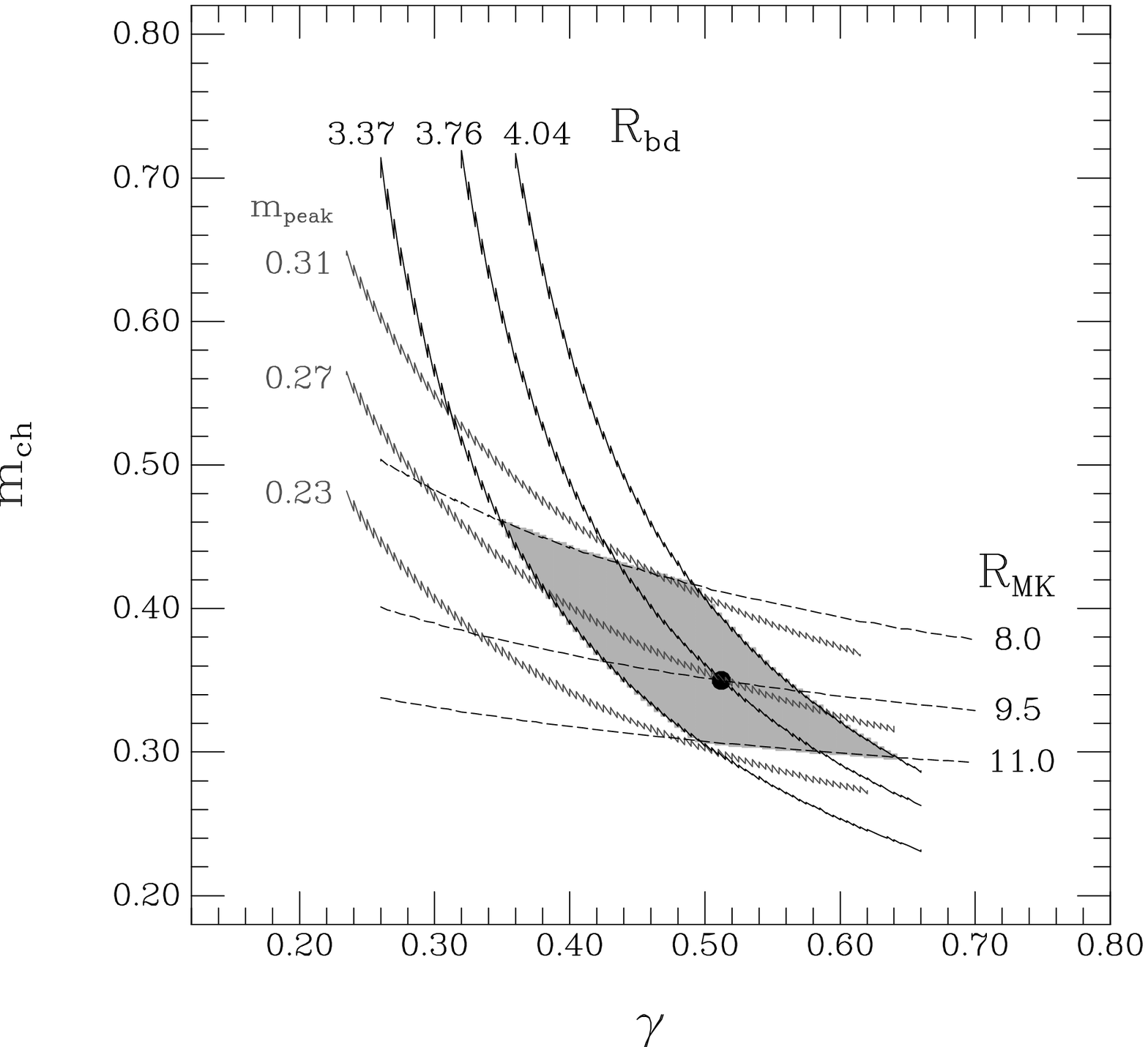}
\caption{The isocurves $R_{MK}=8$, $9.5$ and 11 (dashed); the isocurves $R_\bd^\ind=3.37$, 3.76 and 4.04 (continuous); and the isocurves $m_{\rm peak}= 0.23$, $0.27$ and $0.31$ (grey) for the STPL function with $\Gamma=1.35$. The circle indicates the standard STPL individual-star IMF 
$\psi_{\rm STPL}$, which has parameter values ($\Gamma=1.35$, $\gamma=0.51$ and $\emch=0.35$) that fulfill the constraints $R_\bdh^\ind= 3.76$ and $R_{\MK}=9.5$. 
The errors in the results derived for $\psi_{\rm STPL}$ in Table 2 are the maximum errors when
either $R_\MK^\ind$ or $R_\bd^\ind$ has the best estimate of its value (9.5 and 3.76, respectively).
}
\label{fig:param}
\end{figure}

To determine the best-fit values of
of $\mch$ and $\gamma$, we calculate level curves for $R_\bd^\ind$ and
$R_{\MK}^\ind$ in this parameter space under the assumption that
$\Gamma=1.35$; the values of $m_\ell$ and $m_u$ do not have a significant
effect.  Figure \ref{fig:param} shows 
where $\psi_{\rm STPL}$ fulfills the constraints $R_\bd^\ind= 3.76^{+0.28}_{-0.39}$
 and $R_{\MK}^\ind=9.5\pm 1.5$.
For the standard STPL form of the individual-star IMF, we therefore adopt the values 
$\gamma=0.51$ and $\mch=0.35$, for which $R_\bd^\ind= 3.76$ and $R_{\MK}^\ind=9.5$.
Because of the effects of systematic errors in the determination of both $R_{\MK}$
and $R_\bd$, it is not possible to give 1 sigma error estimates for $\gamma$ and $\mch$.
If we assume that either $R_\MK^\ind$ or $R_\bd^\ind$ has the best estimate
of its value (9.5 and 3.76, respectively) while the other variable has its maximum error, then $\gamma$ lies in the range 0.42-0.58 and
$\mch$ lies in the range 0.30-0.43.
As stated above (see eq. 24), the turnover mass $\mpk$ of the STPL function is in general less than $\mch$. Figure \ref{fig:param} also
shows 
that  $\mpk= 0.27\pm 0.04$.

\section{Comparison and Characterization of IMFs}
\label{sec:comparison}

Table 2 and Figures \ref{fig:ratio-imfs}a,b,c summarize the results for the various IMFs described in \S\S \ref{sec:forms}, \ref{sec:stpl}.
In the table, we separate the results for individual (top) and system or effective
(bottom) IMFs. Since the Salpeter IMF does not include brown dwarfs, we set the lower limit $m_\ell$ on this IMF with $R_{\MK}$. 
The results for the Kroupa, Chabrier, Padoan \& Nordlund, Hennebelle \& Chabrier\footnote{
The Padoan-Nordlund and the Hennebelle-Chabrier theories provide the core mass function 
(CMF)---i.e., the mass function of gravitationally bound regions in a molecular cloud.
Observations show that the CMF is similar to the IMF, but shifted up by a factor
of about 3 (McKee \& Ostriker 2007).
Hennebelle \& Chabrier (2008, 2009) therefore assumed that each core is three 
times the mass of the star(s) that form within it. Padoan et al. (2007) took advantage
of the scale-free nature of self-gravitating simulations to adjust their core masses
directly to the stellar masses. In comparing these IMFs with the observed one, it
is important to note that the observed system IMF is an {\it effective} 
system IMF, as discussed in the Introduction; the difference between the actual
system IMF and the effective one is small compared to the uncertainties in the
relation between the CMF and IMF, however.}
and De Marchi et al
IMFs correspond to 
the parameter values suggested by the authors.
For comparison purposes we estimate that the effective STPL IMF,
$\psi_\stpl^\eff$, is characterized by parameters 
$\gamma=0.57$ and $\mch=0.42$.
These parameters are obtained by assuming that the 
effective peak mass is a factor $1.25$ larger than the peak mass for individual stars (as Chabrier found) and that $R_\bd^\eff= 4.6$.

Figure \ref{fig:ratio-imfs}a  
shows that the Chabrier individual-star IMF 
differs by less than 30\% from the STPL individual-star IMF for $m>0.03$. It is is not surprising
that both IMFs agree for $m>0.1$
since both rely on the RGH02 data, although the STPL IMF also makes use of data on
brown dwarfs that was not avaliable to Chabrier.
If brown dwarfs are omitted from the IMF, then the agreement between the IMFs is even better (<15\%).
Although the Chabrier (2003b) IMF  $\psi_{\rm C03}$ has been updated by Chabrier (2005), we include it in Figure \ref{fig:ratio-imfs}a since recent works continue to use it. 
The Kroupa individual-star IMF differs from $\psi_\stpl^\ind$ by less than a factor 1.6 for $m<1.4$, but since it has 
$\Gamma=1.7$ the difference at large masses increases substantially.
Figure \ref{fig:ratio-imfs}b shows the effective system IMFs. The STPL form of this
IMF is based on Chabrier's (2005) result that the peak mass of the effective
IMF is 1.25 times that for the individual-star IMF, so it is natural that the two
are in good agreement, 
except again for $m<0.03$.
The Kroupa effective IMF differs by less than a factor 1.8 with respect to 
$\psi_\stpl^\eff$ in the entire mass range, but is much larger than all the other IMFs at low masses.
The MPPZ IMF differs significantly at high masses, but it must be borne in mind that the 
high-mass portion of this IMF is based on data that
extend only to 10 \msun; in the mass range $0.04<m<10$, the difference with $\psi_\stpl^\eff$ is less than a factor 1.55.
The Scalo (1986) IMF differs from $\psi_\stpl^\eff$ by less than a factor 2 in the mass range 0.1 - 3.4; note that the difference for $m>3$ would be reduced if it were boosted by $1/b(t_0)$ 
(see Paper II)
and the factors mentioned in \S \ref{sec:Gamma} (Parravano et al. 2003, 2009).
Figure \ref{fig:ratio-imfs}c shows that
the theoretical Padoan-Nordlund IMF agrees well with the STPL IMF (and the Chabrier IMF) for $m>0.05$.
The Hennebelle-Chabrier IMF differs by up to a factor 1.8 from the STPL IMF in this mass range,
but the data comparisons we have made in this paper do
not allow us to determine which is more accurate. Note that 
HC IMF provides an example of an IMF that is flatter at intermediate masses than at
high masses.
It must also be noted that the theoretical IMFs are for single values of the 
parameters, whereas the actual IMF is the result of averaging over a range
of physical conditions.

For each IMF, Table 2 first gives the M to K ratio, $R_\MK$, and 
the brown dwarf ratio, $R_\bd$. The observed values were
used to determine the parameters of the STPL IMF, so it is consistent with
these observations by construction.  
For the individual-star IMFs, the Salpeter IMF 
also agrees with the observed $R_\MK$ by construction,
since its value of $m_\ell$ was chosen to do this;
however, it includes no brown dwarfs.
The Chabrier (2005) IMF is consistent with both the observed
values of both $R_\MK$ and $R_\bd$; the latter agreement is significant, since
$R_\bd$ is based on data published after Chabrier (2005). We did not include the
Chabrier (2003b) IMF in the Table, but we note that it has $R_\MK=11.9$ and
$R_\bd=1.92$, which are far from the observed values.
The Kroupa IMF has a value
of $R_\MK$ that is slightly too high and a value of $R_\bd$ that is significantly too
low. For the effective system IMFs, 
the Chabrier, PN and HC IMFs are all within the errors of the observed $R_\MK$, but only the 
Chabrier IMF is within the errors for the brown dwarf fraction. Note that the MPPZ IMF,
which also has the STPL form, has a value of $R_\MK$ that is far from the observed value.

Next, Table 2 gives
the mean mass of all stars, 
$\avg{m}$, the mean mass of main-sequence
stars (i.e., omitting the brown dwarfs), $\avg{m}_\ms$, and
the peak mass, $m_\pk$. As shown in Table 2, the peak
masses for the PN, HC and Chabrier IMFs are in excellent agreement.
The peak mass for the STPL IMF is somewhat higher than that for the Chabrier
IMF, and that for the MPPZ IMF is somewhat lower. The Kroupa IMF has the lowest
peak mass.
According to the Chabrier (2005) IMF, the average observed mass of an individual
main sequence star is $\avg{m_\ms}=0.69 M_\odot$; the Salpeter IMF (with $m_\ell=0.21$) and
the STPL IMF are consistent with this, but the Kroupa IMF gives a significantly smaller
value, $\avg{m_\ms}=0.43$.  In fact, most of the individual-star quantities 
for the Kroupa IMF in Table 2 disagree with those for all the other IMFs. 
We interpret this as being
due in part to his adoption of a steep slope, $\Gamma=1.7$, for the
high mass ($m > 1$) portion of the individual-star IMF.
As noted above, Ma{\'{\i}}z Apell{\'a}niz (2008) has shown that the correction
for binarity has only a small effect on the slope.

The mass regimes in which disagreements are most significant are at the extremes, as
might be expected. At the low-mass end of the IMF, Table 2 gives
the  fraction of stars that
are born as brown dwarfs, $F_\bd\equiv \dns(<m_\bd)/\dnst$, 
where we adopt $m_\bd=0.075$.
(Recall that we use $m_\bd=0.08$ only for comparison with the observationally
determined $R_\bd$.)
The fraction of stars that are born on the main sequence is $F_\ms=1-F_\bd$.
The brown dwarf number fraction depends slightly on $m_\ell$,
which we assume to be 0.004.  Increasing $m_\ell$ from 0.004 to 0.006 decreases
the fraction of brown dwarfs by about 5\% 
(and the mass fraction by only 1\%).

With the exception of the Kroupa, MPPZ and STPL IMFs, all the IMFs 
rapidly fall off at very low mass, so there is a 
clear distinction in the predictions of the number of low-mass brown dwarfs.
The Chabrier (2005) IMF gives $F_\bd=0.22$ for individual stars, whereas the STPL form
of the IMF implies $F_\bd=0.31$. 
(We note parenthetically that the Chabrier 2003b IMF for individual stars has $F_\bd=0.47$,
which is even larger than the Kroupa value $F_\bd=0.39$;
however, as pointed out above, the Chabrier 2003b IMF is inconsistent with observation.)
The Chabrier (2005) and STPL IMFs differ significantly only for
$m<0.03$, where as yet there is little data. 
For the system IMFs, the Chabrier IMF implies $F_\bd=0.17$.
The MPPZ IMF is in good agreement with this,
but the PN and HC IMFs are
below the Chabrier value by a factor of about 1.6. The STPL IMF has a
value for the system brown dwarf fraction that is a factor of about 1.5 times greater than the
Chabrier value; the Kroupa effective IMF has a brown dwarf fraction more than
twice the
Chabrier value. We note that brown dwarfs have a negligible fraction of the mass;
for example, for $F_\bd\sim 1/3$, the fraction of the mass in brown dwarfs is only
$\sim 2$\%.

\begin{deluxetable}{ccccccccc}
\tablecolumns{9}
\tablewidth{0pc}
\tablecaption{Characterization of IMFs}
\tablehead{
\colhead{IMF$^{\rm ind}$}\tablenotemark{a} 
& \colhead{$R_\MK$} & \colhead{$R_\bd$} &
\colhead{$\langle m \rangle$\tablenotemark{c}}
&\colhead{$\langle m \rangle_{\rm ms}$\tablenotemark{d}}
&\colhead{$m_{\rm peak}$\tablenotemark{e}}
&\colhead{\hspace{-0.2 cm}$F_{\bd}$\tablenotemark{f}}
& \colhead{\hspace{-0.2 cm}$F_{h,\rm ms}/10^{-3}~$\tablenotemark{g}}
& \colhead{$\mu_{h}$\tablenotemark{h} }
 }
 \startdata
$\psi_{\rm Sal}$ &  9.5&---& 0.73& 0.73& 0.21&---&  7.3&  100\\
\tableline
$\psi_{\rm K}$& 11.8&  2.4& 0.28& 0.43& 0.08& 0.39&  2.1&  213\\
\tableline
$\psi_{\rm C}$ &  9.2&  4.0& 0.55& 0.69& 0.20& 0.22&  7.8&   90\\
\tableline
$\psi_{\rm STPL}$ &  9.5$^{+1.5}_{-1.5}$&$ 3.76^{+0.3}_{-0.4}$&$ 0.46^{+0.05}_{-0.04}$&$ 0.65^{+0.08}_{-0.06}$&$ 0.27^{+0.03}_{-0.02}$&$ 0.31^{+0.04}_{-0.04}$&$  6.8^{+1.5}_{-0.9}$&$   97^{+5}_{-7}$\\
\tableline
IMF$^{\rm eff,\, sys}\; \tablenotemark{b}$ & & & & & & & \\
$\psi_{\rm PN}$ &  8.8&  7.1& 0.67& 0.75& 0.24& 0.11&  8.6&   88\\
\tableline
$\psi_{\rm HC}$ &  8.1&  6.8& 0.78& 0.87& 0.25& 0.11& 11.6&   75\\
\tableline
$\psi_{\rm S86}$ & 10.6&---& 0.53& 0.53& 0.29 &---& 1.86&  286\\
\tableline
$\psi_{\rm K}$ & 11.8&  2.4& 0.36& 0.56& 0.08& 0.38&  6.1&   96\\
\tableline
$\psi_{\rm C}$ &  8.0&  5.0& 0.65& 0.77& 0.25& 0.17&  9.2&   85\\
\tableline
$\psi_{\rm MPPZ}$ & 14.4& 5.7& 0.70& 0.82& 0.18& 0.15& 13.9&   59\\
\tableline
$\psi_{\rm STPL}$ & 7.6&$ 4.6$&$ 0.57$&$ 0.75$&$ 0.34$&$ 0.25$&$  8.5$&$ 89$\\
\tableline
\enddata
\tablenotetext{a}{The IMF parameters are:
$\psi_{\rm Sal}(\Gamma=1.35$, $m_{l}=0.21$, $C= 0.167$) for the Salpeter IMF;
see eq. (\ref{eq:kroupa}) with $\gamma_3=-1.7$  and $C= 0.076$ for the Kroupa IMF;
$\psi_{\rm C}(\emch=0.2$, $\sigma=0.55$, $\Gamma=1.35$, $C= 0.315$)
in eqs. (\ref{eq:chabrier}-\ref{eq:chabrier-gt1}) for the Chabrier IMF;
$\psi_{\rm STPL}(\Gamma=1.35, \,
\gamma=0.51,\,  \emch=0.35$, $C= 0.108$) for the STPL with 
the constraints $R_\bd^\ind=3.76$ and $R_\MK^\ind=9.5$,
the quoted errors give the values when 
$R_\MK^\ind$ or $R_\bd^\ind$ has the best estimate
while the other variable has its maximum error.}
\tablenotetext{b}{
System IMF: $\psi_{\rm PN}(\Gamma=1.4$, $\calm = 10$,
$\emch=1$, $C= 0.101$) for the Padoan and Nordlund IMF 
(i.e. $\sigma=1.8$);
$\psi_{\rm HC}(\eta=0.4$, $\calm = 12$,
$\calm_*=\sqrt{2}$, $\emch=1.56$, $C= 0.218$)
for the Hennebelle and Chabrier IMF (i.e. $\sigma=1.9$).
Effective IMFs:
$\psi_{\rm S86}$ from Scalo (1986) Table 7
 for $t_0=12$ Gyr and $b(t_0)=1$.
For the effective Kroupa IMF
see eq. (\ref{eq:kroupa}) with $\gamma_3=-1.3$ and $C= 0.075$;
$\psi_{\rm C}
(\emch=0.25$, $\sigma=0.55$, $\Gamma=1.35$, $C= 0.318$---see
eqs. \ref{eq:chabrier}-\ref{eq:chabrier-gt1})
for effective Chabrier IMF;
$\psi_{\rm MPPZ}
(\Gamma=1.0, \gamma=1.5, \emch=0.15, C= 0.101$)
for the De Marchi et al. IMF;
$\psi_\stpl(\Gamma=1.35, \gamma=0.57, \mch=0.42, C= 0.146)$
for the STPL effective IMF with $m_{\rm peak}=0.34$ and
$R_\bd^\eff=4.6$.
}
\tablenotetext{c} {Average mass of stellar objects 
(main-sequence stars and brown dwarfs) in the IMF.}
\tablenotetext{d}
 {Average mass of main-sequence stars in the IMF.}
\tablenotetext{e} {IMF turnover mass.}
\tablenotetext{f}
 {Fraction of stellar objects that are brown dwarfs 
($0.004<m<0.075$).}
\tablenotetext{g}
{Fraction of main-sequence stars with masses
above $m_{h}=8$, normalized to $10^{-3}$. }
\tablenotetext{h}
{Mass of stellar objects formed per high mass star.}
\end{deluxetable}
\clearpage

For the high-mass end of the IMF, Table 2 gives
two parameters describing stars that
 will explode as core-collapse supernovae, which have masses $m>m_h=8$. In general,
we define $\dnsh\equiv \dns(m>m_h)$, the birthrate of massive stars, and
$F_h\equiv \dnsh/\dnst$, the fraction of all stars that are born with high masses.
Table 2 gives the fraction of main-sequence stars that are born with high mass,
$F_{h,\,\rm ms}\equiv F_h/F_\ms$,
and the mass $\mu _h$ (in solar masses) of stars
formed per high-mass star,
\beq
\mu_h \equiv \frac{\dot M_*}{\edNsh}=\frac{\avg{m}}{F_h},
\eeq
where ${\dot M_*}$ is the total star
formation rate.
The results in Table 2 show that the fraction of individual main-sequence stars that have high masses ($m>8$) is $F_{h,\ms}\simeq (7 - 8) \times 10^{-3}$, and the mass of stars formed per high-mass star is $\mu_h\simeq (90-100) M_\odot$. 
(We do  not include the Kroupa individual-star IMF in this
comparison because of the high value of  $\Gamma$, as discussed above.) 
The fraction of the mass in high-mass stars is about 20\% in each case.
For the system IMFs, 
the Chabrier, PN
and STPL forms all give $F_{h,\ms}\simeq 9\times 10^{-3}$ and $\mu_h\simeq 90 M_\odot$.
The value of $\mu_h$ is smaller for the effective IMF than for the individual-star IMF because
there are more stars with an effective mass above $8\;M_\odot$ than there are individual
stars with such masses; stars with effective masses above $8\;M_\odot$ but actual
masses less than that do not explode as supernovae.
On the other hand, it is possible for one object in the effective IMF to produce more than
one SN, an effect we did not allow for since it depends on the details of binary evolution.
In any case,
the individual-star value of
$\mu_h$ should be used to estimate the rate of core collapse supernovae. (At present,
however, this is just a matter of principle, since the uncertainties in each value are larger than their difference.)
The Kroupa effective system IMF has a similar value of $\mu_h$ but a somewhat lower
value of the number fraction, $F_{h,\ms}$.
The HC IMF gives a somewhat higher fraction of high-mass stars and
a correspondingly lower value of the mass formed per high-mass star.
The MPPZ IMF has a value of the high-mass fraction that is about 1.5 times greater than
the Chabrier value because of its relatively flat high-mass slope, $\Gamma\simeq 1$;
it should be borne in mind that De Marchi et al determined their mass function only for
masses up to $10 M_\odot$.

\begin{deluxetable}{cccccccccc}
\tablecolumns{10}
\tablewidth{0pc}
\tablecaption{Characterization of $\psi_{\rm STPL}$:
 dependence on $\Gamma$}
\tablehead{
\colhead{$\Gamma$} & \colhead{$\gamma$}
& \colhead{$m_{ch}$} & \colhead{$m_{u}$} &
\colhead{$\langle m \rangle$}
&\colhead{$\langle m \rangle_{\rm ms}$}
&\colhead{$m_{\rm peak}$}
&\colhead{\hspace{-0.2 cm}$F_{\bd}$}
& \colhead{\hspace{-0.2 cm}$F_{h,\rm ms}/10^{-3}~$}
& \colhead{$\mu_{h}$}
 }
 \startdata
 1.20& 0.545& 0.315& 120$^{+30}_{-40}$& 0.55$^{+0.01}_{-0.02}$& 0.77$^{+0.01}_{-0.03}$& 0.26& 0.30& 10.6$^{+0.1}_{-0.3}$&   74\\
 1.35& 0.510& 0.350& 120$^{+30}_{-40}$& 0.46$^{+0.01}_{-0.01}$& 0.65$^{+0.01}_{-0.02}$& 0.27& 0.31&  6.8$^{+0.1}_{-0.2}$&   97\\
 1.50& 0.490& 0.380& 120$^{+30}_{-40}$& 0.40$^{+0.00}_{-0.01}$& 0.56$^{+0.00}_{-0.01}$& 0.28& 0.31&  4.4$^{+0.0}_{-0.1}$&  131\\
\enddata
\end{deluxetable}

In order to determine the effect of uncertainties in the high-mass slope, $\Gamma$, and
the upper mass cutoff, $m_u$, we have calculated the various quantities 
characterizing the IMF for the STPL IMF, holding $m_\ell$ and the low-mass constraints 
($R_\MK=9.5$, $R_\bd=3.76$) fixed, but varying the high-mass slope 
($\Gamma=1.2$, 1.35 and 1.5) and the upper mass cutoff ($m_u=80$, 120 and 150).
The upper mass cutoff has little effect on the quantities shown in Table 3. 
In particular, $\mu_{h}=\avg{m}/F_h$ changes less than 1\% with $m_{u}$ because both $\avg{m}$ and $F_h$ increase with $m_{u}$. 
The effects of varying $\Gamma$ are shown in Table 3. The high-mass star fraction
increases by somewhat more than a factor 2 as $\Gamma$ decreases from 1.5 to 1.2;
correspondingly, the mass of stars formed per high-mass star decreases by somewhat
less than a factor 2.

Star formation rates in galaxies are inferred from observations of high-mass stars.
Since the shape of the IMF determines the ratio of the total mass of stars formed to the number
of high-mass stars formed---in other words, $\mu_h$---the shape is essential
for inferring the total star formation rate. It is striking that all the IMFs we
have considered (with the exception of the individual-star Kroupa IMF, which has
an anomalously steep high-mass slope) have a significantly greater fraction of 
high-mass stars, and as a result a significantly lower value of $\mu_h$, 
than does the Scalo (1986) IMF: for an 12 Gyr disk age,
the Scalo IMF gives 
$\mu_h\simeq 290$, about 3 times greater than currently estimated.
Correspondingly, as will be discussed in greater detail in Paper II, the
current estimate for the Galactic star formation rate of $\dot M_*=1.3\pm 0.2\; M_\odot$~yr\e\
(Murray \& Rahman 2010) is several times less than earlier ones (e.g., $4 \;M_\odot$~yr\e, McKee \& Williams 1997), which were based on the Scalo IMF or variations of it.

\section{Observational Tests of the Shape of the IMF}
\label{sec:shape}

\subsection{Comparison with Field Star Mass Function at Low Masses}

\begin{figure}
\epsscale{1.0}
\plottwo{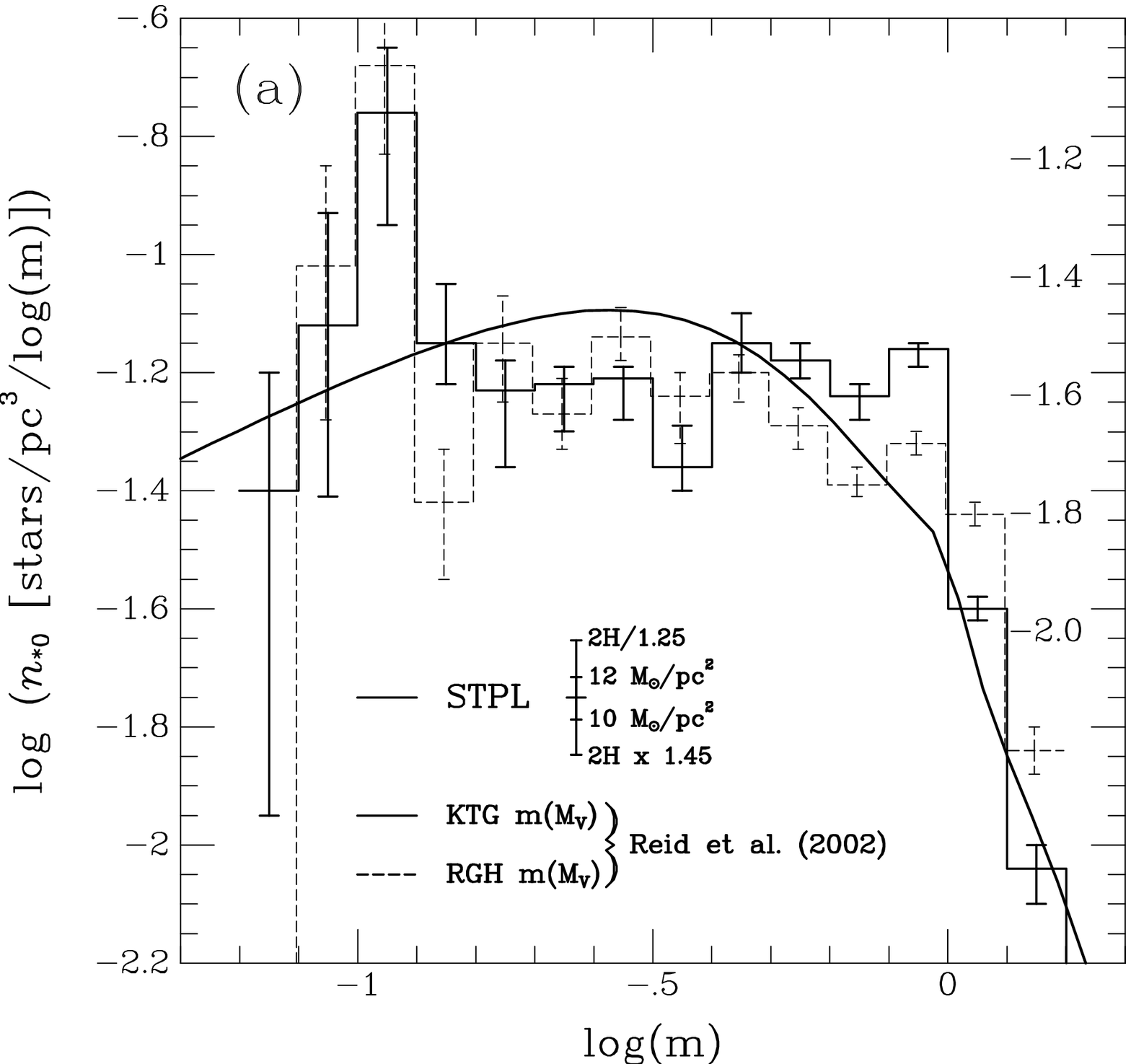}{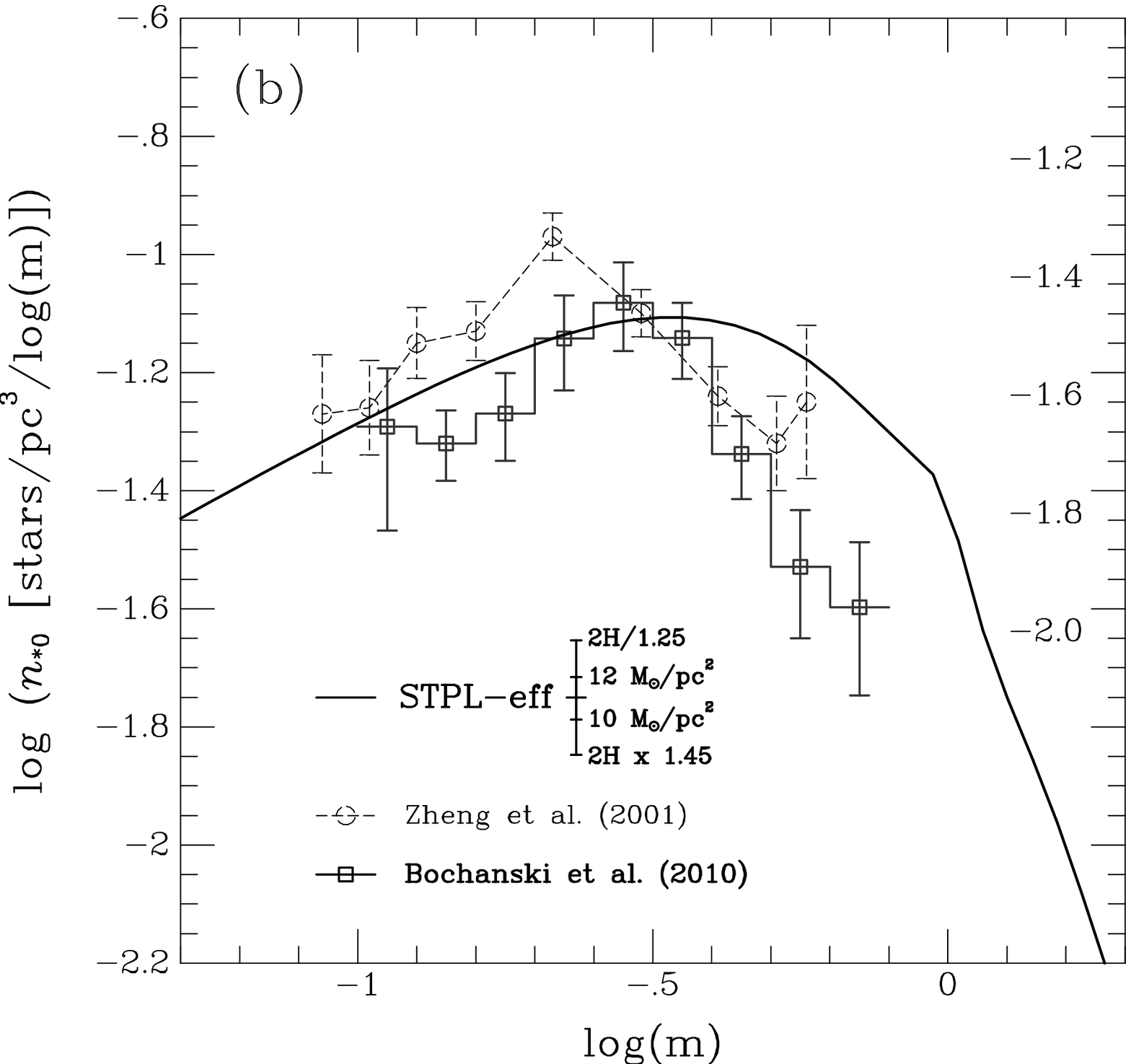}
\caption{
Comparison of the midplane mass function corresponding to the STPL PDMF for a 11 Gyr old disk with 11 \msun\ pc$^{-2}$ in M dwarfs (see text) and the midplane mass distributions derived from field stars observations. 
The error bars beside the STPL label indicates the error from $2H$, assumed to be 25\%, and the uncertainty in the surface density of M dwarfs, taken to be in the range 10 to 12 \msun\ pc$^{-2}$.
The numbers beside the right hand ordinate gives the mass distribution per ln(m) units.
a) The two individual-star mass distributions derived by RGH02 derived from nearby field stars (see text) and the distribution corresponding to the individual-star IMF $\psi_{\rm STPL}$, which has parameter values ($\Gamma=1.35$, $\gamma=0.51$ and $\emch=0.35$).
b) The midplane mass distributions corresponding to the effective IMF $\psi_{\rm STPL}$, which has parameter values ($\Gamma=1.35$, $\gamma=0.57$ and $\emch=0.42$), and the effective mass distributions derived from distant field stars by Zheng et al. (2001) and 
by Bochanski et al. (2010).
}
\label{fig:field-mf}
\end{figure}

The IMF for individual stars, $\psi_{\rm STPL}^\ind$, can be directly compared 
with observed stellar mass functions only for samples in which binary systems are resolved or corrections are made for unresolved systems.
The STPL midplane mass distribution (continuous curve in Fig. \ref{fig:field-mf}a) is derived for a constant SFR during the last 11 Gyr and the scale height $2H(m)$ from Scalo (1986).
The mean SFR during the disk age is adjusted to match a surface mass density of M dwarfs equal to
11 \msun~pc$^{-2}$ (the Flynn et al. (2006) estimate for the thin disk, see Paper II).
As stated in \S \ref{sec:lowmassshape}, most binary systems are resolved in the RGH02 stellar mass function for the local sample of stars.
The histogram in Figure \ref{fig:field-mf}a
shows the RGH02 midplane mass distributions (see their figure 13a,b),
one using the KTG $m(M_{\rm V})$ relation (continuous histogram) and the other using the RGH $m(M_{\rm V})$ relation (dashed histogram).
These mass functions have upticks both at $m\simeq 0.1$
and at $m\simeq 1.0$. We assume that these
apparent features are due to systematic effects such
as uncertainties in the $m(M_V)$ relations 
and the dependence of these relations on metallicity.
From integration of the RGH02-KTG mass distribution 
in Fig. \ref{fig:field-mf}a
and under the assumption that the scale height for M stars is 325 pc (Scalo 1986),
we obtain a surface mass density of M dwarfs
$\Sigma_{{\rm ms}}(<0.63)=10.5$ \msun ~pc$^{-2}$.
This value agrees with the $\sim11 M_\odot$~pc$^{-2}$ estimated by Flynn et al. (2006) for the surface mass density of M dwarfs in the thin disk.

The effective IMF , $\psi_{\rm STPL}^\eff$, is compared in Fig. \ref{fig:field-mf}b with mass functions of distant field unresolved systems.
Observations with the {\it Hubble Space Telescope (HST)} can be used to infer the PDMF of distant field stars.
Zheng et al. (2001, hereafter ZFGBS01) studied main sequence stars with $M_V>8$.
There is no Malmquist 
bias in this study since {\it HST} can see all the Galactic M stars in this direction.  
In addition,  since
M stars have lifetimes substantially in excess of the Hubble time,
the time-averaged IMF and the PDMF (per unit area of disk) are proportional for these stars.
ZFGBS01 considered two color-magnitude relations to infer the
stellar masses. Assuming solar metallicity (method CMR1), they inferred
$\Sigma_{{\rm ms}}(<0.63)=14.3\pm 1.3\;M_\odot$~pc$^{-2}$, which
is in agreement with previous estimates by Gould, Bahcall and Flynn (1996, 1997) using
the same color-magnitude relation. On the other hand, assuming
that the color-magnitude relation is a function of 
height above the Galactic plane due to the vertical gradient 
in metallicity expected in the population of M stars (method CMR2),
they inferred
$\Sigma_{{\rm ms}}(<0.58)=12.2\pm 1.6\;M_\odot$~pc$^{-2}$.
Note that the ZFGBS01 sample includes thick disk stars also and therefore their surface mass density of M dwarfs is expected to exceed the 11 \msun~pc$^{-2}$ value estimated by Flynn et al. (2006) for the thin disk.
We also note that ZFGBS01 presented the results for the CMR1 method
in order to check the consistency with their previous work, but they
regarded the metallicity-gradient model as more physically motivated 
and so more reliable. However, the significant difference
between the PDMFs derived using the CMR1 and CMR2 approximations
illustrates the uncertainties that can be introduced when the
distances to objects are not available and a
color-magnitude relation must be used.
Regarding the effect of binaries,  Gould et al. (1996) showed that the correction for 
binaries has only a small effect on the inferred surface density
$\Sigma_{*,T}$.
The open circles connected by dashed  lines in Figure
\ref{fig:field-mf}b
show the midplane mass distribution from the ZFGBS01 PDMF with the CMR2 method and a scale height $H=325$~pc.
Parravano et al. (2006) fitted the STPL IMF to  the  CMR2 data  and obtained an effective system IMF
characterized by
$\gamma=0.8$ and
$\emch=0.24$;
by comparison, the parameters of the STPL effective IMF found here (which focuses on $R_{MK}$ and $R_{bd}$) are
$\gamma=0.57$ and $\mch=0.42$. The peak mass for the Parravano et al. (2006) IMF
is $0.225 \; M_\odot$, whereas it is $0.34\;M_\odot$ for the one here.

Bochanski et al. (2010) considered a sample of $\sim 15$ million SDSS point sources spread over 8,400 square degrees and obtained an effective mass function  (continuous histogram in Fig. \ref{fig:field-mf}b) in the range $0.1 < m < 0.8$ that is well described by a log-normal distribution with $m_{\peak}=0.25$ and $\sigma=0.28$. This estimate of $m_{\rm peak}$  is in agreement with Chabrier (2005) but, as mentioned in \S \ref{sec:lowmassshape}, $\sigma$ is too small and the ratio $R_{\rm MK}^{\eff}=13.8$ is too high.
The surface mass density of M dwarfs inferred from this mass distribution and a scale height $H=325$ pc
is $\Sigma_{{\rm ms}}(<0.63)=8.1$ \msun ~pc$^{-2}$.
Bochanski et al. (2010) have also estimated the mass function of single stars (not shown in Fig. \ref{fig:field-mf}b) by correcting for unresolved binaries, but their color-magnitude relation has not been corrected for possible metallicity changes with position. This individual-star mass function can be described by a log-normal distribution with $m_{\peak}=0.18$ and $\sigma=0.34$, and the associated surface mass density of M dwarfs is $\Sigma_{{\rm ms}}(<0.63)=10.0$ \msun ~pc$^{-2}$.
In view of the uncertainties in the scale height $H$, the agreement with the Flynn et al. (2006) surface density ($11 M_\odot$ pc\ee) is satisfactory.
These mass distributions can be fitted with a STPL function to infer an individual-star or an effective IMF, as was done in Parravano et al. (2006) for the Zheng et al. (2001, CMR2) data. 
A STPL with $\gamma=0.89$ and $m_{ch}=0.24$ ($m_{peak}=0.23$) fits the constraint $R_{MK}^{eff}=13.8$ from the effective mass function in Bochanski et al. (2010) and $R_{bd}^{eff}=4.6$ from  \S 3.5.
However, in addition to the limitations mentioned in \S \ref{sec:lowmassshape}, Bochanski et al. (2010) consider that the systematic uncertainty in the luminosity function is dominated by the assumed CMR. Additionally we note that Bochanski et al. (2010) considered a CMR that depends on metallicity, but no metallicity correction was applied to the mass-$M_j$ relation; their sample contains stars up to a height of 2 kpc, and such stars have lower metallicity than those in the solar neighborhood (Ivezi\'c et al. 2008).
Figure \ref{fig:field-mf}b shows that the midplane effective mass function derived by Bochanski et al. (2010) and the effective STPL derived here agree within the errors for $m<0.4$, but as the mass increases the Bochanski mass function becomes progresively lower; their mass function is 
Salpeter-like in the mass range 0.32-0.8.

\subsection{Shape of the Brown Dwarf PDMF}

The ideal test to the proposed IMF in the brown dwarf regime is to compare it with the mass distribution of field brown dwarfs, or with the average shape of the mass functions in a representative sample of clusters. Unfortunately, as explained in \S \ref{sec:bd}, at present both approaches have severe difficulties. 
Nevertheless, Allen et al. (2005) estimate that at a 60\% confidence level the 
mass function for objects with masses between 0.04 and 0.10 \msun 
is $d\caln_*/dm \propto m^{-(0.3\pm 0.6)}$, corresponding to $\gamma = 0.7 \pm 0.6$. 
This estimate of $\gamma$ for nearby field brown dwarfs is consistent with
our result for the effective system IMF, $\gamma = 0.57$.
Caballero et al. (2007) studied the young open cluster $\sigma$ Orionis and 
derived its substellar mass function down to 0.006 \msun. The masses were
derived using the theoretical mass-luminosity relation
of Baraffe et al. (2003). The main
result from this study is that the mass function is continuous from
low-mass stars to substellar objects, even below the deuterium-burning mass limit,
with a power-law slope $\gamma=0.4\pm 0.2$ in the mass range $0.006  < m < 0.11 $, in agreement with our standard slope values.

Current data are only now becoming
adequate to measure the shape of the IMF in the brown-dwarf mass
range, and with improving accuracy this measurement will provide a powerful discriminant among
different forms for the IMF. For IMFs that are power laws at low masses, such
as the Kroupa IMF, the MPPZ IMF and the STPL IMF, the fraction of brown
dwarfs below a mass $m$ is (eq. \ref{eq:fbd})
\beq
\frac{F_\bd(<m)}{F_\bd}=\frac{m^\gamma-m_\ell^\gamma}{m_\bd^\gamma-m_\ell^\gamma}.
\eeq
For example, the fraction of brown dwarfs  with $m<0.03\; M_\odot$
(below the mass range 
used to calculate $R_\bd$) is 46\%, 25\% and 52\% of the total number of brown dwarfs
for the Kroupa, MMPZ and STPL individual-star IMFs, respectively (we used $m_\ell=0.004$).
IMFs with a log normal behavior at low masses predict 
smaller values for this ratio than either the Kroupa or STPL IMFs;
for example, the individual-star Chabrier (2005) IMF has 30\% of the brown dwarfs below $0.03\; M_\odot$.

\section{Conclusions}

The stellar initial mass function (IMF) is fundamental to many branches of astronomy.
Observations suggest that when averaged over long times and large areas of the Galactic disk,
it is universal in shape (Bastian et al. 2010). We argue
that  it should also be simple in shape, since there are no known processes that could
imprint features on the IMF given the wide range of conditions under which stars form.
We propose that the IMF can be represented by a smoothed two-power law (STPL),
$$
\psi_{\rm STPL}(m) = C m^{-\Gamma} 
	   \left\{1-\exp\left[{-(m/\emch)^{\gamma +
	   \Gamma}}\right]\right\}~~~~~(m_\ell<m<m_u).
$$
This IMF has five parameters: The lower-mass cutoff, $m_\ell$, which we set equal
to the minimum fragment mass expected on the basis of opacity
effects, $0.004\;M_\odot$ (Low \& Lynden-Bell 1976; Whitworth et al. 2007);
the upper mass cutoff, $m_u$, which appears to be about $120 M_\odot$ for
individual stars (\S \ref{sec:mu}); the high-mass slope parameter, $\Gamma$, which we
take to be the Salpeter value of 1.35 (\S \ref{sec:Gamma}); the low-mass slope,
$\gamma$; and the characteristic mass, $\mch$, which is comparable to the mass
at which the IMF peaks, $m_\peak$. In order to determine these latter two
quantities, we use integral properties of the IMF that are set by observation:
(1) the ratio $R_{\MK}$ of the number of stars (mainly M-dwarfs) in the range
$0.1-0.6 M_\odot$ to the number 
of stars (mainly K dwarfs) in the range $0.6-0.8 M_\odot$, as inferred from
the Reid et al. (2002; RGH02) sample of nearby stars (\S \ref{sec:lowmassshape}); 
and (2) the ratio $R_\bd$ 
of the number of stars in the mass range $0.08 - 1 M_\odot$ 
divided by the number of objects (mostly brown dwarfs) between $0.03$ and $0.08 M_\odot$
(\S \ref{sec:bd}). The RGH02 data are for individual stars, but the brown dwarf data
must be corrected for unresolved binaries.
With the observed values of $R_\MK$ and $R_\bd$, we infer
$\gamma=0.51$ and
$\emch=0.35$ for the STPL individual-star IMF; the corresponding peak mass
is $m_{peak}=0.27$.
For these values of the parameters, the normalization factor is $C=0.108$.
The STPL form for the effective system IMF is found by using the value
of $R_\bd$ found for systems in \S \ref{sec:bd} and assuming that the peak mass for systems is a factor $1.25$ larger than the peak mass for individual stars, as Chabrier (2005) found.
The resulting parameter values are
$\gamma=0.57$ and
$\emch=0.42$; the corresponding peak mass and normalization factor are respectively $m_{peak}=0.34$ and $C=0.146$.

A STPL form for stellar mass functions was first proposed by Paresce \& De Marchi
(2000), who applied it to globular clusters. De Marchi \& Paresce (2001) showed that
this form applied to young Galactic clusters as well, and drew the important
conclusion that
the characteristic mass of Galactic clusters increases with the age of the cluster.
Our approach in studying the mass function is complementary to theirs: They 
have focused on analyzing individual clusters, whereas we have used data
on nearby stars to infer the shape of the IMF for low-mass main sequence stars
and a broad range of data to infer the high-mass slope of the IMF. We have also
considered a broader mass range, using data from $0.03\;M_\odot$ to over
$100\; M_\odot$, whereas they have focused on the mass range between
$0.1\; M_\odot$ and $10\; M_\odot$.  For their average effective mass function,
they find $\Gamma\simeq 1$, $\gamma\simeq 1.5$, and $\mch=0.15$;
the resulting values for $R_\MK$ (14.4) and $R_\bd$ (5.7) are not consistent with
the values we have determined.
The reasons for this discrepancy are not entirely clear. We note that
the parameter $\gamma$ determines the number of substellar objects; we use observations of the ratio of the number
of substellar objects to the number of low-mass stars ($R_\bd$) to infer $\gamma$, whereas De Marchi et al. (2010) do not include objects less than 0.1 in their analysis.

The STPL IMF proposed here agrees well with the Chabrier (2005) IMF
(i.e. to within 30\% for $m>0.03$).
Both are based on the RGH02 data, but the STPL IMF also makes use
of data on brown dwarfs that was not available in 2005. 
As shown in Table 2, the mean stellar mass is $\avg{m}\simeq 0.5$; if
only main sequence stars are included, the mean mass is $\avg{m}_\ms\simeq 0.7$.
This agreement
applies to both the individual-star and effective system IMFs; the latter
is not surprising, since we used Chabrier's (2005) result on the ratio of
the peak masses in converting the individual-star IMF to the effective
system IMF. Chabrier's earlier version of the individual-star IMF
(Chabrier 2003b) is inconsistent with both the value of $R_\MK$ determined
from local stars and with the value of $R_\bd$ determined from
very young clusters.

As shown in figure \ref{fig:ratio-imfs}a, the 
Salpeter form of the individual-star IMF, which
had its lower mass limit $m_\ell$ adjusted to fit the observed value of $R_{MK}$,
is above the STPL IMF at all masses since its normalization factor ($C=0.167$) is 1.55 times the corresponding STPL value. The Salpeter and STPL IMFs agree within 30\% for $m>0.4$ when the Salpeter IMF is reduced by a factor 1.55 or when the IMFs are normalized to produce a unit mass.
At large masses there are significant disagreements between the Kroupa individual-star IMF
and the others since  Kroupa's high mass slope ($\Gamma=1.7$) is substantially larger than others, but for $m<1.4$ it agrees  to within a factor 1.6 with the STPL IMF.

Figure \ref{fig:ratio-imfs}c shows the theoretical system IMFs.
Both the Padoan \& Nordlund (2002) 
and Hennebelle \& Chabrier (2008, 2009) forms for the IMF are in reasonable agreement (within 30\%)
with the STPL and Chabrier IMFs for $m>0.05$, except for $m>1$ where Hennebelle \& Chabrier IMF exceed STPL IMF up to a factor 1.8. They are both within the errors of
the measured value of $R_\MK$, but they underpredict the number
of brown dwarfs ($R_\bd$ is too high). The success of these theoretical
forms for the IMF in matching the data is impressive, particularly in view
of the fact that the actual IMF is the result of stars forming in a wide
range of physical conditions, whereas the theoretical values are for
only one set of conditions.

The mass range in which the disagreement among the IMFs is greatest is
the brown dwarf regime, which is also the mass range in which the data
are most limited. The STPL, MPPZ and Kroupa forms for the effective IMF
assume that the very low mass range is described by a power law, whereas
Chabrier assumes a log normal for that mass range. Both theoretical
IMFs (PN and HC) have a log-normal form built in from the log normal
form for the density spectrum found in numerical simulations.
A sensitive measure of the IMF in the brown dwarf regime is the fraction
of brown dwarfs predicted to be below some mass; for example,
under the assumption that brown dwarfs extend down to $0.004\; M_\odot$,
the STPL and Kroupa forms for the IMF predict that about half of all brown dwarfs have
masses below $0.03\;M_\odot$, whereas the Chabrier form
predicts 
that only about 27\% of brown dwarfs have such a low mass;
PN and HC forms predict even lower values, 14\% and 15\% respectively.
Since the formation of brown dwarfs requires unusual conditions,
the IMF of brown dwarfs is a powerful test of star formation
theories (McKee \& Ostriker 2007).

The IMFs considered here agree reasonably well on the
proportion of high-mass stars ($m>m_h=8$), which will
produce core-collapse supernovae. For the individual-star IMF, the mass
of stars formed per high-mass star is $\mu_h\simeq 90-100$.
(The effective system IMF has a somewhat lower value of
$\mu_h$ because it includes stars of mass $m<8$ in binaries with
effective masses above $8\; M_\odot$.) This value of $\mu_h$ is
significantly smaller than that for the Scalo (1986) IMF,
which has $\mu_h=280$. As will be discussed in greater detail
in Paper II, the result is that current estimates of the star formation 
rate in the Galaxy are several times lower than those in previous decades.

We acknowledge support from the NASA Astrophysical Theory Program in
RTOP
344-04-10-02, which funded The Center for Star Formation Studies, a
consortium of
researchers at NASA Ames, University of California at Berkeley, and
University of California at Santa Cruz.
AP was supported as a Senior Associate
for part of this research by the National Research Council and by
the Universidad de Los Andes by CDCHT project C-1653-09-05-B.
The research of CFM was supported in part by NSF grant AST-0098365 and by
the Groupement d'Int\'er\^et Scientifique
(GIS) ``Physique des deux infinis (P2I)."

\appendix
\section{Binary Correction}

Recall from \S 1 that there are three forms of the IMF, and correspondingly three
ways to count stellar and substellar systems: $\caln_*^\ind$ is the number
of individual stars, counting two stars for each binary, three for each triple,
etc; $\caln^\sys$ is the number of stellar and substellar systems in which
binaries, etc., are counted as single objects and characterized by their
total mass (or primary mass); and $\caln^\eff$ is the number of stellar
and substellar systems in which each system is characterized by an effective
mass, $\meff$, determined from unresolved observations in which multiple systems
are assumed to be single. In the text, we use two observed ratios to fix
the form of the IMF: $R_{MK}$, which is roughly the ratio of M stars to K stars,
and $R_\bdh$, which is the ratio of the number of low-mass stars to
readily observable brown dwarfs (actually, stars in the mass range $0.03-0.08\; M_\odot$).
Current data allow one
to determine $R_{MK}$ for individual stars, but the data for
$R_\bdh$ do not resolve binaries; the data therefore determine
$R_\bdh^\eff$. To relate the observed $R_\bdh^\eff$ to the required $R_\bdh^\ind$,
we first need to account for binaries and then allow for the difference between
the effective mass and the actual mass. 

We denote the mass range for low-mass stars  ($0.08<m<1 $)
by the subscript ``$\ell$" and the mass range for readily observable brown
dwarfs ($0.03<m<0.08$) by the subscript ``$\bdh$."
For example, $\nle$ is the number of low-mass stellar systems with effective
masses in the low-mass range and $\nbdhe$ is the number of
substellar systems with effective masses in the range
$0.03-0.08 M_\odot$.
The observed ratio $R_\bdh$ is based on infrared observations that
did not resolve binaries,
\beq
R_\bdh=R_\bdh^\eff \equiv\frac{\caln_*^\eff(0.08\leq\meff\leq 1)}{\caln_*^\eff
(0.03\leq\meff\leq 0.08)}
\equiv\frac{\nle}{\nbdhe},
\eeq
whereas we need 
\beq
R_\bdh^\ind\equiv\frac{\nli}{\nbdhi}
\eeq
to fix the form of  the IMF.

\subsection{Ratio of Low-Mass Stars to Low-Mass Systems, 
$\caln_{*\ell}^\ind/\caln_{*\ell}^\sys$}

First we consider the ratio of the number of individual low-mass stars to
the number of low-mass systems. 
We characterize binaries
by the mass of the primary, $m_1$. The subscript
``$b$" denotes binary stellar systems; thus,
$\caln_\bl$ is the number of binary stellar systems
in which
both the primary and the secondary
stars are in the low-mass range.
The number of stellar systems in which the mass of the star, for single systems,
or the mass of both the primary and
the secondary, for binary systems, is in the low-mass range is denoted
$\caln_{*\ell}^\sys$.  
(Thus, systems are placed in the low-mass category based on the individual stellar
masses, not on the total mass of the system.)
The binary fraction for low-mass stellar systems is then $f_\bl=\caln_\bl/\caln_{*\ell}^\sys$.
Note that brown-dwarf (BD) companions of low-mass stars are not counted
in $f_\bl$; these will be discussed below.
For the moment, let us neglect the number of low-mass companions of stars with masses exceeding
$1\;M_\odot$ (which we shall term
``supersolar stars"), we have
\beq
\caln_{*\ell}^\ind\simeq (1+f_\bl)\caln_{*\ell}^\sys.
\label{eq:nstl}
\eeq
According to Lada (2006) the PDMF of locally observed stars has a single
star fraction of 2/3. Since most of the stars in the PDMF are less than
$1\;M_\odot$, this  corresponds to a single star fraction below $1\; \:M_\odot$ of
2/3, or a binary fraction $f_\bl\simeq 1/3$. 

	We now determine the correction for low-mass
companions of supersolar stars. For a flat distribution
of secondary to primary mass (Mazeh et al. 2003), one can show that about half the 
supersolar binaries
have a low-mass companion. We obtain an upper limit
on the effect of these low-mass companions by assuming a binary fraction
of unity for supersolar stars; then the number of low-mass companions
of supersolar stars is 1/2 the number of supersolar systems, and the number of supersolar
stars is 3/2 times the number of supersolar systems.
The ratio of low-mass companions of supersolar stars (denoted by the subscript
``ss") to low-mass stars
is then about 
\beq
\frac 12 \left(\frac{\caln_{*\rm ss}^\sys}{\caln_{*\ell}^\ind}\right)
\simeq \frac 12 \left(\frac{\frac 23 \caln_{*\rm ss}^\ind}{\caln_{*\ell}^\ind}\right)
\simeq 0.05,
\eeq
where we obtained the ratio of the number of supersolar stars to low-mass stars
in the individual-star IMF by iteration. Thus,
the number of low-mass stars in equation (\ref{eq:nstl}) should
be increased by about 5\%. We conclude that
\beq
\caln_{*\ell}^\ind\simeq 1.05\times \frac 43 \; \caln_{*\ell}^\sys = 1.40 \caln_{*\ell}^\sys.
\eeq

\subsection{Correction for the Effective Mass, $\caln_{*\ell}^\sys/\caln_{*\ell}^\eff$}

Next, we must estimate the ratio of the number of low-mass stellar systems,
$\caln_{*\ell}^\sys$, to the number of stellar systems with low effective masses,
$\caln_{*\ell}^\eff$. Fortunately, this correction too is very small. We approximate the luminosity in a photometric
band as a power-law in mass,
\beq
L_\pb\propto m^\jpb.
\label{eq:lpb}
\eeq
The results of Delfosse et al. (2000) show that $\jpb\simeq 2$ 
for $0.1<m<0.7$ for both
the J-band and the K-band (their more complicated fits are needed to
extend the fits over a slightly larger mass range). For a low-mass binary with
a mass ratio $q=m_2/m_1$, where $m_2$ is the mass of the secondary,
the luminosity of the system corresponds to an effective mass
\beq
\meff=m_1(1+q^2)^{1/2}.
\label{eq:meff}
\eeq
We estimate the effect of the effective mass by assuming that all
the binaries have the typical value of $q$. For low-mass
systems, this is about $\frac 12$  according to Mazeh
et al. (2003), corresponding to $m_1=0.89 \, \meff$. 
The effective number
of low-mass stellar systems, $\caln_{*\ell}^\eff$,
omits the binaries in the mass range $0.89-1\; M_\odot$ and
includes the binaries in the mass range just below $0.08\; M_\odot$.
We obtain an upper limit on the ratio $\caln_{*\ell}^\sys/\caln_{*\ell}^\eff$ by omitting
the latter:
\beq
\caln_{*\ell}^\eff\simeq\caln_{*\ell}^\sys-\caln_b(0.89\leq m_1\leq 1).
\eeq
The binary fraction for solar mass stars was inferred to be 0.57
by Duquennoy \& Mayor (1991). 
(The Lada (2006) binary fraction used above for $m<1$ is smaller because it 
includes less massive stars.)
Using equation (\ref{eq:nstl}), we
then obtain
\beqa
\caln_{*\ell}^\eff&\simeq&\caln_{*\ell}^\sys\left[ 1-\frac{0.57\caln_*^\sys(0.89\leq m_1\leq 1)}
{\caln_{*\ell}^\sys}\right],\\
&\simeq&\caln_{*\ell}^\sys\left[ 1- \frac{0.57(1+f_\bl)}{1.57}\;
\frac{\caln_*^\ind(0.89\leq m_1\leq 1)}{\caln_{*\ell}^\ind}\right].
\eeqa
The determination of the ratio
$\caln_*^\ind(0.89\leq m_1\leq 1)/\caln_{*\ell}^\ind$ must be done iteratively;
we find that it is about 0.03 
so that the correction in going from the effective mass to the primary mass 
results in an increase of at most about 1\%, implying $\caln_{*\ell}^\eff\simeq \caln_{*\ell}^\sys$.
We conclude that $\caln_{*\ell}^\ind\simeq 1.4 \caln_{*\ell}^\eff$.

\subsection{Effect of Brown Dwarfs in Binaries, $\caln_\bdh^\ind/\caln_\bdh^\eff$}

	We turn now to the effect of brown dwarfs in binaries, which enters the
denominator of $R_\bdh$. Very low-mass (VLM) binaries are defined as those
with primary masses less than $0.1\; M_\odot$. Reid et al. (2006)
and Burgasser et al. (2007) find VLM binary fractions of $24^{+6}_{-2}\%$ and
$22^{+8}_{-4}\%$, respectively. 
Since the binary fraction tends to decrease as the
mass decreases (Burgasser et al. 2007), 
we adopt a binary fraction for
BD-BD binaries of 20\%, recognizing that this is uncertain.
Since BD-BD binaries tend to have similar masses (Burgasser et al. 2007),
we assume that all the BD secondaries are in the target mass range of
$0.03-0.08\; M_\odot$.

	BD secondaries of H-burning stars are comparable in
number to the BD secondaries of BD primaries.
Burgasser et al. (2007) summarize the data on BD companions of
low-mass stars. Imaging surveys indicate that the number of BD
secondaries is about 1/3 that of hydrogen-burning secondaries; radial
velocity surveys, which are sensitive to smaller separations, give a somewhat
smaller fraction. We adopt a value $f_\bl/3$ for the fraction of 
low-mass systems with BD secondaries, recognizing that this too is
uncertain. Then the total number of brown dwarfs is the number of
systems with BD primaries, plus the number of BD companions of
BD primaries, plus the number of BD companions of H-burning stars:
\beq
\caln_\bdh^\ind=\caln_\bdh^\sys + \caln_{\bdh,\,b}+\frac 13 f_\bl\caln_{*\ell}^\sys.
\eeq
Keep in mind that in order to conform to the convention adopted by
Andersen et al. (2008), we are including stars with masses between the
H-burning limit $\simeq 0.075\;M_\odot$ and $0.08\;M_\odot$ in
$\caln_\bdh$. 
We have just seen that $\caln_{\bdh,\, b}\simeq 0.2\caln_\bdh^\sys$, so
we can solve this equation to find
\beq
R_\bdh^\ind=\frac{(\caln_{*\ell}^\ind/\caln_{*\ell}^\sys) R_\bdh^\sys}{1.2+\frac 13 f_\bl R_\bdh^\sys}\simeq \frac{1.4 R_\bdh^\sys}{1.2+\frac 19 R_\bdh^\sys},
\label{eq:rbdhind}
\eeq
since the binary fraction of low-mass stars is $f_\bl\simeq \frac 13$ (Lada 2006) and
we previously found that the total number of low-mass stars is
about 1.4 times the number of low-mass systems.

In principle,
there could be a correction in going from the system counts to
the effective system counts (i.e., from $\caln_\bdh^\sys$ to
$\caln_\bdh^\eff$), but as in the case of the low-mass stars, we expect
this correction to be small. The simple approach adopted for estimating
this effect for low-mass stars (starting with eq. \ref{eq:lpb}) does
not work for brown dwarfs, since spectroscopic information is also
used in estimating the mass of the brown dwarf; furthermore, for field brown
dwarfs, the estimate of the mass is coupled to an estimate of the age.
Since this correction is small, and we earlier found that 
$\caln_{*\ell}^\sys\simeq\caln_{*\ell}^\eff$, we have $R_\bdh^\sys\simeq R_\bdh^{eff}$.
According to the estimation in \S \ref{sec:bd} $R_\bdh^\eff=4.6^{+0.5}_{-0.65}$.
Equation (\ref{eq:rbdhind}) then implies $R_\bdh^\ind\simeq 3.76^{+0.3}_{-0.4}$.

{}

\end{document}